\documentclass[10pt,conference]{IEEEtran}
\IEEEoverridecommandlockouts

\usepackage{cite}
\usepackage{float}
\usepackage{graphicx}
\usepackage{hyperref}
\usepackage{booktabs}
\usepackage[tight]{subfigure}
\usepackage{caption}

\usepackage{afterpage}
\usepackage{graphicx}

\usepackage{amsmath,amssymb,amsfonts}
\usepackage{algorithmic}
\usepackage{textcomp}
\usepackage{xcolor}
\usepackage{tkz-euclide}
\usepackage{framed}
\usepackage{tikz}
\usepackage{adjustbox}
\usepackage{url}
\usepackage{libertine}
\usepackage{pdflscape}
\usepackage{enumitem}

\usetikzlibrary{positioning, arrows, shapes,chains}
\usepackage{color}
\usepackage{transparent}
\graphicspath{img/}
\usepackage{cleveref}

\Crefformat{figure}{Figure~#1}
\Crefmultiformat{section}{Sections~#2#1#3}{ and~#2#1#3}{, #2#1#3}{ and~#2#1#3}

\def\BibTeX{{\rm B\kern-.05em{\sc i\kern-.025em b}\kern-.08em
    T\kern-.1667em\lower.7ex\hbox{E}\kern-.125emX}}

\begin{document}

\title{The TechDebt Game - Enabling Discussions about Technical Debt}

\author{\IEEEauthorblockN{Anonymous Authors}}

\author{

\IEEEauthorblockN{Marion Wiese\IEEEauthorrefmark{1},\\Angelina Heinrichs, \\Nino Rusieshvili}
\IEEEauthorblockA{\textit{Department of Informatics,} \\
\textit{University of Hamburg}\\
Hamburg, Germany 
    \\\IEEEauthorrefmark{1}marion.wiese@uni-hamburg.de}
\and

% \IEEEauthorblockN{Angelina Heinrichs}
% \IEEEauthorblockA{\textit{Department of Informatics,} \\
% \textit{University of Hamburg}\\
% Hamburg, Germany}
% \and

% \IEEEauthorblockN{3\textsuperscript{rd} Nino Rusieshvili}
% \IEEEauthorblockA{\textit{Department of Informatics,} \\
% \textit{University of Hamburg}\\
% Hamburg, Germany \\}
%
\and

%\IEEEauthorblockN{4\textsuperscript{th} Luka Zdravkovic}
%\IEEEauthorblockA{\textit{Department of Informatics,} \\
%\textit{University of Hamburg}\\
%Hamburg, Germany \\}
%\and

\IEEEauthorblockN{Rodrigo Rebouças de Almeida}
\IEEEauthorblockA{\textit{Federal University of Paraíba} \\
\textit{Department of Exact Sciences} \\
Rio Tinto/PB, Brazil \\
rodrigor@dcx.ufpb.br}
\and

\IEEEauthorblockN{ Klara Borowa}
\IEEEauthorblockA{\textit{Warsaw University of Technology,}\\ 
{Institute of Control and}\\
{Computation Engineering}\\
Warsaw, Poland \\
klara.borowa@pw.edu.pl}

}

% plus authors in add. mat: 
% author = {Wiese, Marion and Heinrichs, Angelina and Rusieshvili, Nino and Rebou{\c{c}}as de Almeida, Rodrigo,  Klara Borowa},

% \and
% \IEEEauthorblockN{6\textsuperscript{th} Klara Borowa}
% \IEEEauthorblockA{\textit{dept. name of organization (of Aff.)} \\
% \textit{name of organization (of Aff.)}\\
% City, Country \\
% email address or ORCID}
%}

\maketitle

\begin{abstract}
    % Context: The importance of the research questions addressed by the review. 
    % 1. A general statement introducing the broad research area of the particular topic being investigated.
    \textit{Context.} 
    Technical Debt (TD), defined as software constructs that are beneficial in the short term but may hinder future change, is a frequently used term in software development practice. Nevertheless, practitioners do not always fully understand its definition and, in particular, conceptual model. Previous research highlights that communication about TD is challenging, especially with non-technical stakeholders. Discussions on this topic often cause conflicts due to misunderstandings related to other stakeholders’ perspectives.
    % Objectives: The questions addressed by the systematic review. 
    % 2. An explanation of the specific problem (difficulty, obstacle, challenge) to be solved.
    % 3. A review of existing or standard solutions to this problem and their limitations. 
    \textit{Goal.} 
    We designed a board game to emulate TD concepts to make them tangible to all stakeholders, including non-technical ones. The game aims to encourage discussions about TD in an emulated and safe environment, thereby avoiding real-life conflicts. 
    % Methods:	Data Sources, Study selection, Quality Assessment and Data extraction. 
    \textit{Method.} 
    To evaluate the game’s effectiveness, we surveyed 46 practitioners from diverse domains, positions, and experience levels who played the game in 13 sessions following extensive testing during its development.
    In addition to the players' general feedback, we examined situations where players recognized new insights about TD or connected game scenarios to real-life experiences. 
    % Results: Main finding including any meta-analysis results and sensitivity analyses. 
    % 4. An outline of the proposed new solution. 
    \textit{Results.} Overall, the feedback on the game and its enjoyment factor were highly positive. While developers and software architects often connected game situations to their real-world experiences, non-technical stakeholders, such as scrum masters, product owners, and less experienced developers, encountered multiple new insights on TD.
    Numerous players have shifted their attitudes toward TD and have outlined a plan to modify their behavior regarding TD management.
    %instances where they learned something new about TD.
    % Conclusions: Implications for practice and future research.
    % 5. A summary of how the solution was evaluated and what the outcomes of the evaluation were.
    \textit{Conclusions.} 
    Although the game may not lead to long-term behavior change among stakeholders, participants' feedback provides evidence that it might serve as a valuable starting point for team discussions on technical debt management.
    %The game has the potential to bring all stakeholders to a similar level of knowledge to discuss TD and TD management.
    
\end{abstract}

\begin{IEEEkeywords}
Technical Debt, Technical Debt Management, Game-Based Learning
\end{IEEEkeywords}

%\linenumbers

    \section{INTRODUCTION}
    \label{sec:Introduction}
        
        %introducing TD and ATD

            Avgeriou et al. defined Technical debt (TD) as a result of a trade-off decision for IT constructs between short-term beneficence and long-term maintainability~\cite{Avgeriou2016a, Kruchten2019}. 
            These constructs can occur in various elements of the system or software engineering process, e.g., code, tests, or requirements~\cite{Ernst2021}. 
            There is a substantial body of work on how TD and its related concepts should be modeled, named, and described~\cite{Li2015, Kruchten2019, Ernst2021}, e.g., a single instance of TD is named a TD item~\cite{Kruchten2019}.
            Avgeriou et al.presented an overall conceptual model of TD~\cite {Avgeriou2016a}, which has been used as a basis for various research, e.g.,~\cite{rios2018tertiary, Junior2022, wiese_it_2023}. 
            As part of the InsighTD project~\cite{InsighTD2022, Rios2020, Ramac2021}, the main causes and consequences of TD have been extensively described.

            %The tertiary studies on TD management by Rios et al.~\cite{rios2018tertiary}, and by Junior et al.~\cite{Junior2022} both provide extensions to the conceptual model regarding TD activities, TD types and strategies \& technologies.
            %These days research tends to focus, in particular, on architectural TD (ATD)~\cite{Martini2015a, Besker2018b, Verdecchia2021c}. 
            %ATD was identified as being one of the most dangerous types of TD~\cite{Martini2015a}, because there is a lack of refactoring strategies that would enable ATD repayment after it is incurred~\cite{Besker2018b}.
        
        %Rationale
            Initially, W. Cunningham employed the term Technical debt (TD) as a metaphor to explain to business stakeholders how fast delivery of makeshift software, results in the need for code rewrites~\cite{Cunningham1992} . 
            Communicating the TD concepts to business stakeholders is still challenging, as current research shows~\cite{Verdecchia2021, Besker2022, wiese_it_2023, avgeriou_technical_2023}.
            The communication on the TD topic poses the risk of misunderstandings and unproductive discussions between different stakeholders~\cite{Wiese2022}.
            Moreover, the vision paper on TD management highlights the relevance of socio-technical aspects of TD~\cite{avgeriou_technical_2023}.
            Hence, this study's \textbf{rationale} is to enable and improve communication between the stakeholders involved in creating IT systems and harmonize their knowledge level about TD.
            %By this, we strive to promote informed discussions on embedding TD management into the software development life cycle between all stakeholders.
        
        %Goal +
            For this purpose, we designed a board game in an offline and online version to imitate the TD concepts in an emulated environment with the \textbf{goal} of enabling safe and engaging communication free of real-life conflicts between stakeholders of various backgrounds, including technical and business stakeholders.
            The game was designed along what we called \textit{aha}-moments, i.e., recognition moments where players could identify new insights about TD or relate game scenarios to their real-life experiences.
            We intended the game as a starting point to discuss TD rationally and, ultimately, set up a TD management process within a team or company.
 
        %Method
            A total of 46 stakeholders from diverse domains, experience levels, and positions played the TechDebt Game in 13 game sessions of usually four stakeholders each. 
            The games were played in Brazil, Poland and Germany. 
            We evaluated the outcome by surveying the players after each game, asking for feedback, determining their motivation to change their behavior, and specifically, assessing the \textit{aha}-moments.
    
        % RQs
            Our \textbf{research questions (RQ)} focus on evaluating the effectiveness of using a board game to teach TD concepts, stimulate discussions among practitioners, and provide an enjoyable and psychologically safe environment for exploring and discussing technical debt.
            \begin{itemize} [leftmargin=*, itemsep=0pt]
                
                \item [] 
                \textbf{RQ 1: To what extent does the game emulate the real-life experiences that technical stakeholders encounter?}
                To be meaningful to all stakeholders, the game must provide realistic elements to provide genuine insights while remaining abstract enough to ensure an engaging flow.
                Thus, we asked the players if they had experienced situations related to the game and how enjoyable they found playing it.
                
                \item [] 
                \textbf{RQ 2: Do the players encounter new insights and learn about additional facets of TD management?}
                Since one of the game's goals is to educate stakeholders about various TD-related concepts to harmonize the level of knowledge between stakeholders, we questioned the players about which \textit{aha}-moments were previously unfamiliar to them and how these offered new perspectives on TD. 
                
                \item [] 
                \textbf{RQ 3: How does the game change players' attitudes and possibly behaviors regarding TD?}
                %Are the players open to changing their behavior and discussing TD management after the game?}
                Finally, the game is set up to stimulate discussions on embedding TD management into the player's real-world workflows, e.g., their software development life cycle. 
                Thus, we asked about the players' attitudes towards TD and plans to change their behavior regarding TD management after they played the game.
            \end{itemize}
        
        %Contribution - general
            As the \textbf{contributions} of our study, we provided a strategic board game built on the theoretical foundation of game-based learning that emulates the real-life experiences of developers regarding TD. 
            We demonstrated that the TechDebt Game can be used to educate stakeholders, foster communication on TD management, and support discussion about TD management strategies.
            Particularly, business stakeholders and junior developers profited from new insight.
            We revealed that the game might encourage a player's behavior change, which provides a starting point for embedding TD management if played in a team working together on one IT system.
            
        % study outline
            
            In the following section, we compare our study with related work (\Cref{sec:RelatedWork}). 
            We present the theoretical background on game-based learning that forms the basis for understanding the game design in~\Cref{sec:background} and the game design itself in~\Cref{sec:Game}.
            We explain the game's evaluation approach in~\Cref{sec:Method}, and the respective results in \Cref{sec:Results}. 
            Afterward, we discuss our findings in~\Cref{sec:Discussion} and the threats to the validity of our study in~\Cref{sec:ThreatsToValidity}.
            With~\Cref{sec:Conclusion}, we conclude our paper, summarize the contributions for practitioners and researchers, and identify future work.
            The additional material comprises the TechDebt Game and details on the evaluation~\cite{AdditionalMaterial}. The TechDebt Game is also available online~\cite{TechDebtGame}.
        
    \section{RELATED WORK}
    \label{sec:RelatedWork}
        Our related work section focuses on game-based learning approaches used in software engineering and, particularly, TD management.
        
        %There is a variety of games that aim to promote understanding of important software development topics through game-based learning. 

        Problems and Programmers is a card game on the subject of software development designed by Baker et al.~\cite{baker_problems_2003}. 
        %CUT: It represents an emulation of the software process in various phases. 
        %The aim is to complete the fictitious software project first.
        %The competitive design allows players to observe other people's strategies and compare them with their own. 
        %In addition to the learning objective of teaching the players about the software development process, there is also the goal of creating a clear, understandable, interactive game with realistic feedback for the players.
        Due to the small number of participants, the findings only indicate that, according to the players' subjective opinions, the game facilitates collaborative learning.
        %CUT: The authors concede that a compromise must be made between the scope of learning aspects about software processes and a functional game.

        PlayScrum and GetKanban v4.0 teach frameworks for agile software development. %~\cite{fernandes_playscrum_2010}. % and the concepts and practices associated with it. 
        %The game emulates the Scrum process and lets the players take on the role of a Scrum Master leading a software development project. 
        %It is played with a game board, product backlog cards, problem cards, concept cards, developer cards, artifact cards, and dice. It is designed for two to five players. 
        Fernandes et al. evaluated PlayScrum with 13 master's students using questionnaires~\cite{fernandes_playscrum_2010}. 
        The players rated it as entertaining and effective for learning Scrum concepts. 
        %However, the study leaders noted that it was difficult for players without knowledge of Scrum to understand the game and its mechanics. 
        %Therefore, they recommended using the game as a supplement during the study and not in isolation for it to be effective. 
        %It was also recommended that the game's maps be further developed and digitized
        %GetKanban v4.0 is a collaborative board game for learning the agile software development method ``Kanban'' 
        GetKanban v4.0  was evaluated against predefined learning objectives~\cite{heikkila_teaching_2016}. %The aim of the game is to generate as much financial value as possible by producing new features in a software system. To achieve this, the players have to make effective resource management decisions in various roles.
        %GetKanban v4.0 has been simplified compared to its predecessor version. Many game elements that had additional rules or complicated mechanics were removed, and the changes were generally intended to provide a simpler, faster, and smoother game experience. 
       % CUT: Heikkilä et al.~\cite{heikkila_teaching_2016} tested and evaluated the game against predefined learning objectives. 
        Heikkilä et al. found that the game was motivating but had no significant effect on the knowledge transfer of Kanban methods.

        DecidArch (v2) and SmartDecisions are board games that teach decision-making in software architecture.
        %It is an enhanced version of the previous gameDecidArch[Lag+18]. It pursues the same. 
        DecidArch learning objectives are (1)~the rationale of design decisions, (2)~understanding the diversity of solutions, and (3)~the impact of changing design decisions~\cite{DeBoer2019}. 
        The game was evaluated with students in two game iterations to improve the game design.
        SmartDecisions focuses on the ``Attribute-Driven Design'' method (see~\cite{bass_software_2021}) and was evaluated with 41 participants, including students, educators, and practitioners, employing feedback forms and game results~\cite{cervantes_smart_2016}.
        %The game results showed no differences between experienced and inexperienced players.
        %De Boer et al.~\cite{DeBoer2019} evaluated DecidArch in two game iterations with 83 students in 22 groups in the first and 77 students in 20 groups in the second iteration to improve the game design.
        %De Boer et al.~\cite{DeBoer2019} evaluated DecidArch utilizing 83 students in 22 groups. 
        %It proved to be effective in teaching the learning objectives, although the third learning objective (dependencies) was too dependent on chance. Accordingly, the game was revised, and DecidArch v2 was developed to address the weaknesses that had previously been identified and to support the three learning objectives better.
        %The DecidArch v2 study involved 
        %CUT: They evaluated the first version with 83 students in 22 groups and the second with 77 students in 20 groups.
        %CUT: For both versions of the game, participants completed the same questionnaire. %surveys, which consisted of Likert scale questions and open-ended questions. The second evaluation also showed the effectiveness of the game, but some remaining problems were identified. The third learning objective (dependencies) was better supported, but problems remained in supporting the second learning objective (differences) and the third learning objective. 
        %De Boer et al. recognized the need for improvement regarding limited playtime, unclear relationships between design options, and the game's emphasis on individual design decisions.
        
        Hard Choices is the only empirically evaluated board game designed to teach the concept of TD~\cite{Ganesh2014}. % in the software development process. %CUT candidate.
        In this game, players compete against each other to be the first to bring a product to market. %CUT:, earning points as they develop the software. 
        Players can skip some development phases by incurring TD in the form of bridge cards evaluated as minus points in the final calculation. 
        Ganesh et al. tested the game with 41 IT students using pre- and post-tests and a survey. 
        The results revealed a significant improvement in the understanding of TD after the game. 
        %Ganesh was critical of the lack of a control group to better validate the instructional methods of the game and the small sample size of the teachers. She recommended testing the game several times with different teachers to rule out bias.

        Hard Choices is the only game similar to our game as it focuses on TD concepts. 
        However, the main purpose of this game and its evaluation is to educate IT students.
        Our game is designed to foster not only learning but discussion among and behavior change in practitioners, including non-technical stakeholders, and was evaluated using players from practice. 
        %CUT candidate:
        %Furthermore, the game mechanics are distinct: our game incorporates the concept of building a system and completing tickets, while Hard Choices uses dice to move across a path, following actions and consequences along the way.

        Other educational games on the topic of TD, such as the ``Technical Debt Game''~\cite{TDGame1} and the ``Technical Debt Game-for non-technical people''~\cite{TDGame2} were not evaluated, and thus, their effectiveness has not been determined. 
       % However, no studies were publicly available on these games, and thus, their effectiveness cannot be compared.

    \section{THEORETICAL BACKGROUND}
    \label{sec:background}

    In this section, we provide the theoretical background on game-based learning and game design that forms the basis for understanding the design approach of our game.

        \subsection{Game-based Learning}

        Game-based learning is a learning method in which teaching content is passed on to the learners by playing a game. 
        Suits et al. define a game as an interactive activity to achieve a goal, which is restricted by rules~\cite{suits_what_1967}. 
        In educational games, the game aims to impart knowledge through the activity of playing. 
        %This involves restructuring the actual learning task into a game result~\cite{plass_foundations_2015}. 
        Some game definitions provide greater detail regarding the presence of a conflict,
%        This refers either to the conflict modeled by the game with the game itself or the conflict of competing with other players. 
 %       Both types of conflict 
        which plays a prominent role in cooperative and competitive games~\cite{pivec_aspects_2003, plass_foundations_2015}.

        In their article on the basic laws of game-based learning, Plass et al. defined the four most fundamental arguments for using game-based learning~\cite{plass_foundations_2015}: %~\cite{shi_game_2015}.
            \textit{Motivation} of players has been proven to be increased by game-based learning compared to learners participating in conventional learning activities~\cite{huizenga_teacher_2017}. %, which again improves the learning outcomes~\cite{shi_game_2015}. 
            The increased motivation has, in turn, improved the learning outcomes~\cite{shi_game_2015}. 
            %A combination of intrinsic and extrinsic motivation is crucial~\cite{garris_games_2002, sousa_leadership_2019}. 
            \textit{Player engagement} is the player's willingness to engage with the game and its content. 
            %CUT: This engagement can arise from different engagement categories, including cognitive engagement, affective engagement, behavioral engagement, and sociocultural engagement. 
            %Increased learning outcomes only occur in the presence of cognitive engagement; all other types of engagement only help if they promote cognitive engagement.
            \textit{Adaptability} describes the flexible adjustment of the difficulty to the player's abilities at the game's runtime. 
            %Assistance such as additional challenges can be used to regulate the level of difficulty in the game. 
            \textit{Graceful failure} is a frequently occurring advantage of Game-based learning and describes the concept of consequence-free failure.
            Consequence-free refers to the fact that all effects of failure do not leave the virtual environment of the game. 
            This allows players to experiment with the game, e.g., exploring a style of play that contrasts with their real-life approach. %receive both positive and negative feedback and learn from it. 
            %Mistakes do not leave the context and are not transferred to reality, where damage could occur.
            %This allows players to explore a style of play that contrasts with their real-life approach, letting them learn from the outcomes.

        \subsection{Game Design}

        \textit{\textbf{Educational game elements.}}
        While not every game leads to improved learning outcomes, there are some game elements that can improve outcomes regarding educational content~\cite{garris_games_2002}. 
        %Some game elements particularly support improved learning outcomes~\cite{garris_games_2002}. 
        Thus, it is integral to defining the individual game element categories to transfer real-life knowledge. 
        Plass et al. define the following game element categories~\cite{plass_foundations_2015}: % include game mechanics, aesthetic design, narrative design, incentive systems, musical value, content, and skills~\cite{plass_foundations_2015}. 
        %Depending on the game's genre, the extent to which the individual game elements appear varies. 
        %In some genres, elements can be omitted entirely.
        \begin{itemize}[leftmargin=*, itemsep=0pt]
            \item \textit{Game mechanics} are the player's interactions with the game, including every playful activity in which the game's rules are followed. % is included in this category. 
            %These interactions usually take place during the course of the game and, therefore, play an important role in teaching content. 
            Game mechanics can either teach content and skills directly or provide feedback for other actions. 
            %An example of a game mechanic could be that players should reach their goal by laying out cards. 
            %If players play a card that contradicts the learning content, the game can pass on feedback to the player by deducting points.
            \item \textit{Aesthetic design} describes the presentation of information and game concepts. %, including the handling of the individual game elements. 
            %It is, therefore, important for elements of aesthetic design that crucial information is repeated or emphasized. 
            An appealing and, at the same time, informative visual design is relevant to all game elements, including cards, game boards, or even coins.
            \item \textit{Narrative design} is crucial for creating the virtual environment.
            %Digital games or emulations can use captivating dialogs and stories to increase player motivation. 
            In board games, texts can tell a story and embellish the virtual environment. 
            Narratives are an important medium for passing on theoretical knowledge.
            %Good narrative design and, thus, a detailed environment increase the intrinsic motivation of players~\cite{shi_game_2015}.
            \item \textit{Incentive systems} are essential for increasing player motivation. 
            %Incentive systems are differentiated according to the motivation they are trying to achieve. 
            Intrinsic motivation can be achieved by promoting enjoyment and, thus, increasing the desire to play the game. 
            Extrinsic motivation can be reached by fostering competition outside of the specific game environment, e.g., by awarding points for a final tally.
            \item The \textit{musical value} of a game refers to the management of attention and mood through the use of music, background noise, or auditory cues. 
            \item \textit{Content and skills} formulate the learning objectives that shall be acquired with the game. %Either theoretical content or skills in the subject matter are conveyed. 
            Theoretical content and skills in a subject form the basis for good game elements.
            %The learning objectives can differ, for example, when consolidating already known knowledge instead of preparing for future learning. 
        \end{itemize}

        \textit{\textbf{Creation of an educational game.}}
        %Now that there is a general understanding of technical debt and the possible game elements have been listed, the following section will clarify how this understanding can be translated into individual elements. To this end, 
        Pivec et al. outlined the following steps for creating educational games~\cite{pivec_aspects_2003}:
        
%        \begin{enumerate}
%\item 
            (1)~\textit{Learning approach.} 
            Before starting the development process, the theoretical foundation of the learning objectives must be defined. 
            %It must be clarified how game players should cognitively engage with the material. 
            The content and the approach to the previously described game element category ``content and skills'' are determined. % in the context of the game elements.
 %           \item 
            (2)~\textit{Creating a model world.}
            The medium and genre of the game are determined, which has a direct impact on the other game elements. 
            The specification of the model world provides the framework for the rest of the game. 
  %          \item
            (3)~\textit{Executing the details.} 
            The theoretical foundations and the model world are brought together through narrative design. 
            The first elements of knowledge transfer can be incorporated. % into the context of the game. 
   %         \item 
            (4)~\textit{Incorporating learning support structures.} 
            %The theoretical learning approach can now be put into practice in this model world.
            Clear goals that the player should achieve and appropriate consequences for pursuing these goals are defined. 
            %Appropriate consequences for pursuing these goals must be defined. 
            %Learning support structures guide players into the subject matter and strengthen the skills to be learned.
    %        \item 
            (5)~\textit{Transforming learning activities into interactions with the game interface.} 
            %When transforming the learning activities into interactions with the game, the previously described processes for achieving the learning objective are realized. 
            The game mechanics, e.g., drawing cards, rolling dice, or moving pieces, for solving the game's underlying conflict are defined. %described. 
            %If players interact with the game by drawing cards, rolling dice, or moving pieces, these actions are part of the game mechanics.
     %       \item 
            (6)~\textit{Transforming concepts into interface objects.} 
            %Until this step, the game is purely conceptual. 
            %However, when concepts are transferred to interface objects, 
            The perceptible objects of the game are created, including visual representations of the game board, cards, or game resources (aesthetic design), background music (musical value), and points and achievements (incentive system). 
      %  \end{enumerate}

    \section{TECHDEBT GAME}
    \label{sec:Game}
\begin{figure}%[H]
        \centering
        \includegraphics[width=0.45\textwidth]{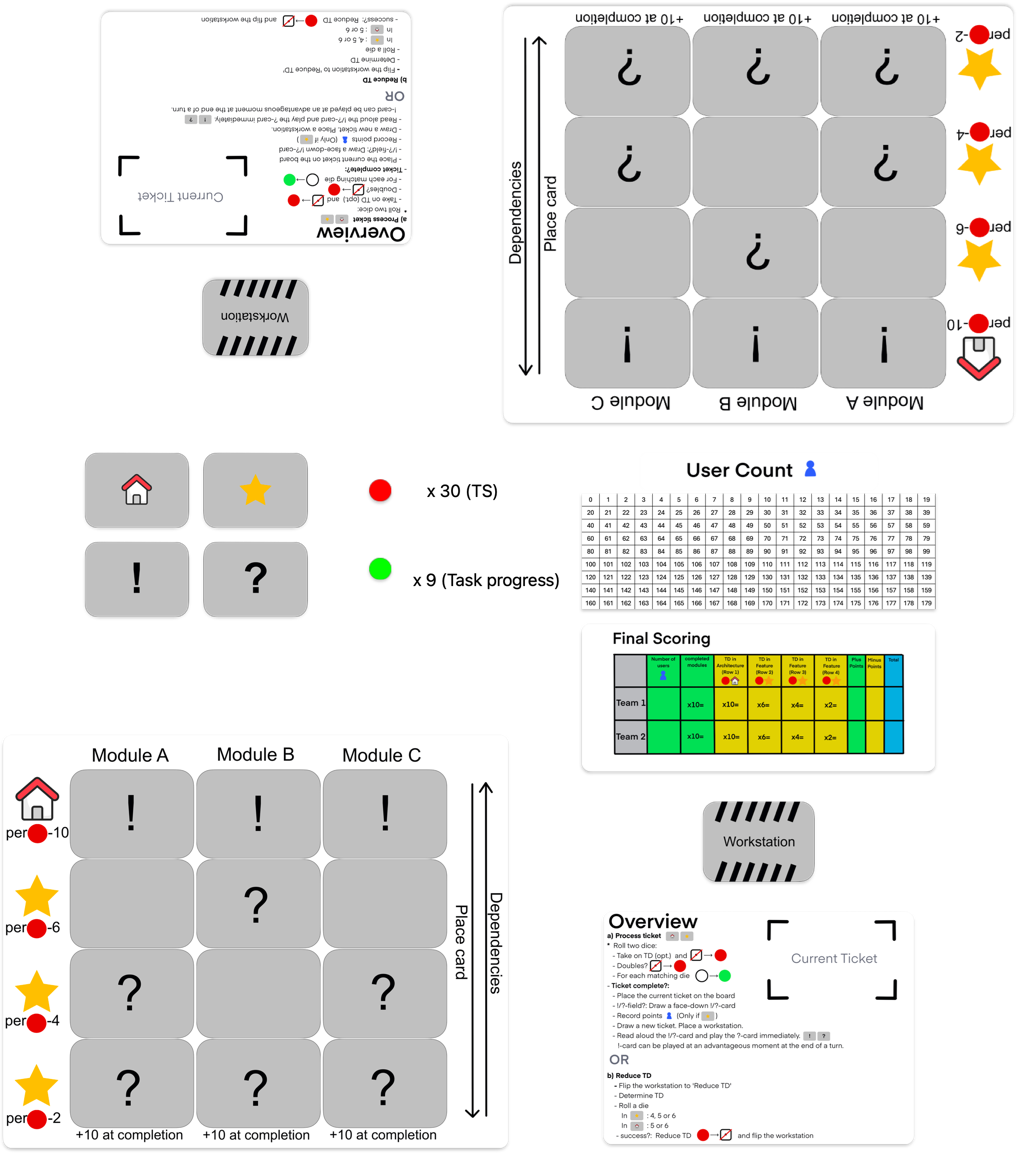}
        \caption{Overview of the Game Board (details see~\cite{AdditionalMaterial}}
        \label{fig:game_plane}
        %\vspace{-3mm}
    \end{figure}

    The TechDebt Game is a board game that aims to playfully increase understanding and discussion of TD by emulating the software development process. 
    Two teams of two players compete against each other for a predefined duration (e.g., 60 minutes) to develop a system by figuratively working on software architecture and features.
    As the team completes tasks, they accumulate users (the game's points).
    Although taking on technical debt may speed up task completion in the short term, it ultimately makes future tasks more difficult to accomplish and results in a lowered score by the end of the game.
    The team that accumulates the most users (points) within the duration of the game wins.
    Players can employ different strategies to manage TD and encounter the respective results (e.g., ``gracefully failure'').
   The two-on-two format encourages discussion among team members, creating a collaborative dialogue where all participants gain from exchanging and learning from each other's insights. % ; and the predefined duration creates time pressure that motivates the incurrence of technical debt.
    %Figure 1
    \Cref{fig:game_plane} shows the TechDebt Game board.
    The game is available as a PDF and must be augmented with dice and game pieces. 
    The game PDF and manual are part of the additional material~\cite{AdditionalMaterial}.
    
        \subsection{Game design, elements and mechanics}
\begin{figure}%[H]
            \centering
            \includegraphics[width=0.45\textwidth]{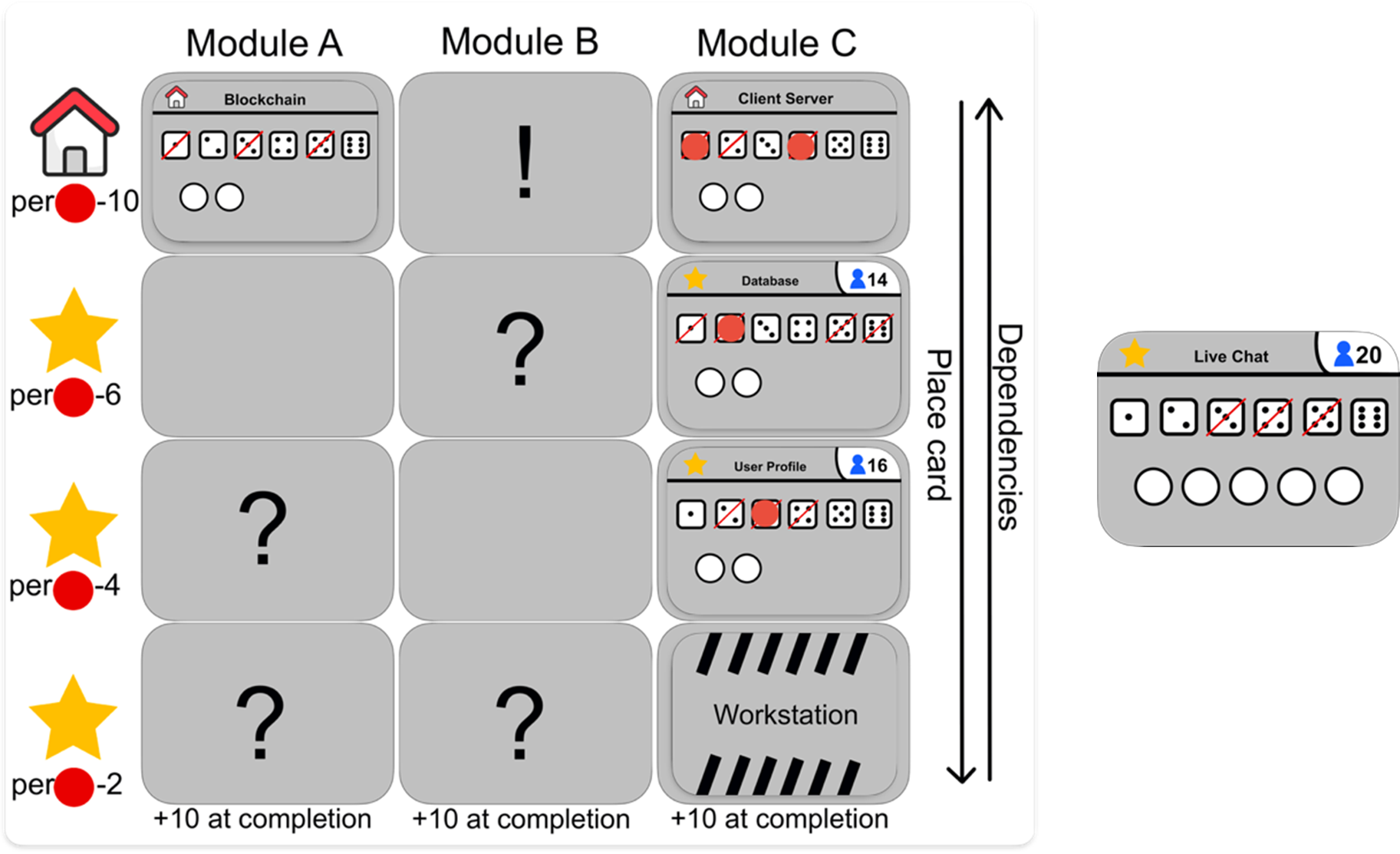}
            \caption{Example for dependencies created by incurring TD (red tiles) -- only rolling a six will complete tasks}
            \label{fig:dependencies}
            %\vspace{-3mm}
        \end{figure}

        \begin{itemize} [leftmargin=*, itemsep=0pt]
            \item \textit{Game board:} 
            The game board represents the progress of a project and the placement of the modules. 
            Each team builds one system consisting of three modules (A-C).
            \item \textit{Architecture tickets and feature tickets:} 
            To start a module, an architecture ticket has to be completed and placed in the first row of the system.
            After that, completed feature tickets can be placed on the subsequent rows.
            Feature tickets earn points, which are represented by the users gained through a feature.
            Each ticket shows the six digits of a dice, some of which are blocked, i.e., crossed out by a red line on the ticket's respective dice symbol (see \Cref{fig:dependencies}).
            Two dice are rolled to ``work'' on a ticket. 
            If one of the rolled digits is not blocked, progress is made, i.e., one task of the ticket is completed. 
            The number of tasks required to complete the ticket is shown with empty circles.  
            \item \textit{TD incurrence:}
            TD can be incurred consciously by adding TD to a dice blocked for the ticket, resulting in immediate success in completing one task if that digit was rolled.
            TD can be incurred unconsciously by throwing a double. 
            Then, TD must be incurred at the respective digit, and two tasks will be completed immediately.
            \item \textit{Workstation/TD Repayment card:}
            This card is placed at the location where the completed ticket will be laid out to identify dependencies on other tickets.
            To repay TD, the card has to be flipped over to the TD repayment side, and the team has to dedicate a turn specifically for the repayment.
            To repay TD on feature tickets, a four, five, or six must be rolled.
            As repaying architectural TD is harder, only rolling a five or six leads to a successful repayment. 
            This means they are investing time that could have been spent on another task, representing the cost of addressing and repaying the TD.
            \item \textit{TD dependencies:}
            The current ticket depends on all previous tickets in the same module, which means that TD from previous tickets hinders the current ticket's development. 
            Each digit that contains a TD on one of the depending tickets is blocked when working on the current ticket, i.e., if the current ticket would allow a roll of six to be successful, but a previous ticket contains a TD on a six, the six is no longer available to complete a task.
            \Cref{fig:dependencies} shows an example where the tasks of the last ticket can only be completed by rolling a six.
            \item \textit{Event and action cards:} 
            Event cards (question marks) have immediate negative effects and represent the causes and consequences of TD. 
            Action cards (exclamation marks) can be used strategically and represent positive TD management actions. %measures that help manage TD.
        \end{itemize}

        \subsection{Game Design Iterations}
        We designed the game in three main iterations, each focusing on one topic: (1) set up the initial game design, (2) simplify the game by removing non-TD-related game elements like the scrum cycle, (3) digitalizing the game and adapting the board game so both fit the requirements of digitalization.  
        For iterations 1 and 2, we played nine games overall. 
        During the second iteration, we optimized the game along pre-defined learning objectives, which we derived from current literature on TD, e.g.,~\cite{Kruchten2019, Avgeriou2016a, wiese_it_2023}.
        We referred to those learning objectives as \textit{aha}-moments, a term that reflects the expression people often use when they recognize or understand something new. 
        %This highlights that the goal is not just to acquire knowledge but also to recognize and connect real-life situations, encouraging discussions around them.
        \Cref{tab:ahamoments} gives an overview of all \textit{aha}-moments and their descriptions.
        During the second iteration, each game was recorded, transcribed, and coded with the \textit{aha}-moments.
        The \textit{aha}-moments were then counted to ensure they could be encountered by the players.
        
        Furthermore, small changes were made throughout all iterations and sometimes between games, optimizing the game flow, the rules, and the introduction. 
        For example, we simplified the game process by removing the prioritization between various feature tickets. 
        %CUT candidate:
        %Instead, the features are now distributed randomly. 
        While this prioritization is a real-life task that happens regularly as part of the software development life cycle, it prolonged the game without changing the player's perception of TD.

        % Another example regards the incurrence of TD. 
        % While initially TD could be incurred consciously only before throwing the dice, players can now incur TD afterward. 
        % Incurring TD is much more tempting if one already knows which die numbers were thrown.
        % This elevates not only the chance of a faster solution but is instantly a faster solution. 
        % This change made the game much more realistic and improved the overall game experience.

	\begin{table*}
    	    \centering
    	    \footnotesize
    		\caption{\textit{Aha}-moments and their explanations \footnotesize (causes and consequences are taken from Wiese et al.~\cite{wiese_it_2023})}
                \label{tab:ahamoments}
            \scalebox{0.9}{
        	\begin{tabular} {lll} %
                            %{p{1cm}p{1cm}p{4cm}}
    			\hline
        	   	Group & Variable  & Description\\ 
    			\hline 
                        Causes			& Time			& Time pressure can be a cause of TD (deadlines).																					\\
                        				& Budget		& Cost pressure can be a cause for TD (license costs).																			    \\
                        				& Business		& Business decisions can be a cause of TD (change in requirements, change in strategy).												\\					
                        				& Management	& Management decisions can be a cause of TD (broken communication, poorly planned projects).											\\							
                        				& Personnel		& Personnel can be a cause of TD (lack of personnel, inexperienced or unmotivated personnel, frequent changes).\\		
                        				& Technology	& chosen technology can be a cause of TD (outdated technology).																		\\
                        				& Decisions		& Incorrect decisions can be a cause of TD (architectural decisions).																\\		
                        				& Awareness		& Lack of awareness of TD can be a cause of TD.																			            \\
                        				& Chains		& Causes of TD can trigger other causes of TD.																			            \\
    			\hline 
                        Incurrence		& Conscious		& TD can be incurred consciously.																									\\		
                        			& Unconscious	& TD can be incurred unconsciously.																									\\	
    			\hline 	
                        Consequences	& Time			& TD can lead to more time expenditure (overtime, missed deadlines, longer development process).										\\								
                        				& Budget		& TD can lead to higher costs (optimization costs, project becomes more expensive).													\\					
                        				& Business		& TD can negatively affect the business (unmet requirements, loss of customers, legal consequences).							\\	
                        				& Management	& TD can negatively affect management (lack of controllability, future risk).													\\					
                        				& Personnel		& TD can lead to personnel problems (terminations, stress, new developers having to be trained).										\\								
                        				& Technology	& TD can lead to technology problems (maintainability, bugs, dead-end).																\\		
                        				& Chains		& Consequences of TD can trigger further consequences of TD.																			\\
    			\hline 
                        Vicious		    & Inner			& TD can lead to further TD (broken window phenomenon).																			        \\
                        Cycle			& Outer			& Consequences of TD can become causes for new TD.																			        \\
    			\hline 
                        Repayment		& Difficult		& Paying back TD is difficult.																			                            \\
                        				& Time-consuming &	Repaying TD is time-consuming.																			                        \\
                        				& Benefits		& The repayment of TD can create advantages for further development.																	\\	
                        				& Simplified	& Certain measures make it easier to repay TD (refactoring, engaging specialists, communication).									\\	
    			\hline 
                        Architecture  	& Critical		& TD items in architecture are the most critical debts.																			        \\
                            		& Hard to repay	& TD items in architecture are the hardest to repay.																			            \\
                        				& Prevents TD		& Architecture can help deal with TD.																			                \\
    			\hline 
                        TD 			    & Identifying TD	& To fix TD, they must first be detected.																			                \\
                        management		& Prioritizing TD 	& To make decisions, TD must be prioritized.																			        \\
                        			& Ignoring TD	& It is not always reasonable to fix (all) TD.																			            \\
    			\hline 
                        Business		& Invisible		& TD are invisible in themselves and can only be recognized through symptoms.														\\				
                        			& Perspective	& Causes and consequences of TD can be difficult to discern from a business/management perspective.									\\	
                    \hline    		
                \end{tabular}
                }
    	\end{table*}

        \subsection{Summary}
        We consider that the TechDebt Game meets the key features of an educational game~\cite{plass_foundations_2015} as described in~\Cref{sec:background}:
        \begin{itemize} [leftmargin=*, itemsep=0pt]
            \item \textit{Game mechanics:} Activities like rolling dice (``working''), placing TD tiles, integrating tickets into the system, or drawing action and event cards form the game's basis. %%placing game pieces for tasks,  and recording points.
            \item \textit{Aesthetic design:} 
            The game features a minimalist design emphasizing key functionalities by using specific colors and symbols, such as red tiles for TD or a house to represent architecture. %red tiles for TD.
            \item \textit{Narrative design:} 
            The game follows a straightforward narrative: 
            A system comprising features and software architecture has to be built, and whoever develops a better system with more users wins. 
            The predefined duration adds time pressure that motivates the competition and might lead to an accumulation of technical debt.
            Using TD in the system can speed up the development of the current ticket but increases the difficulty of completing dependent tickets. 
            %complicates the completion of dependent tickets.
            %TD in the system can make the development of the current ticket faster but the completion of dependent tickets more difficult.
            \item \textit{Incentive system: }
            Points and visible progress in the game encourage competition with the opposing team. 
            Action and event cards act as rewards and punishments.
            \item \textit{Musical value:} 
            As this was originally a board game, there is no background music. 
            Additionally, music could have drowned out the discussion between players.
            % We assume that the benefit of using music in this setting would be very limited.
            \item \textit{Content and skills:} 
            The learning objectives, i.e., \textit{aha}-moments shown in ~\Cref{tab:ahamoments}, served as guiding principles for the game's development. %  was optimized towards in each iteration. 
            %A list of all \textit{aha}-moments can be derived from the additional material~\cite{AdditionalMaterial}.
        \end{itemize}

        The game is available as a PDF file as part of the additional material~\cite{AdditionalMaterial} and for future development on GitHub~\cite{TechDebtGame_Tabletop}. It must be augmented with materials like dice, or game stones. 
        A digital version is available online~\cite{TechDebtGame_Digital}.
        
    \section{METHOD}
    \label{sec:Method}
      In this section, we present how we evaluated the TechDebt Game by playing it with IT stakeholders and surveying them.
        
	\begin{table}
    	    \centering
    	    \footnotesize
    		\caption{Participants \footnotesize (PM/PO - project manager/ product owner)}
                \label{tab:participants}
            \scalebox{0.8}{
        	\begin{tabular}{lllllll} % p{2,1cm}
    			\hline
        	   	Game & loc- &Coun- &  Domain  & Position & Experience \\ % & TD-\\
        	   	   & ation&   try  &   Domain &          &  (in yrs.) \\ % & Term\\
    			\hline 

                      1 & on-site	&GER & Electronics 		& Developer	& 6 - 10 			\\ % & y \\
                      1 & on-site	&GER & Electronics 		& IT-Manager& 3 - 5 			\\ % & y \\
                      1 & on-site	&GER & Electronics 		& Developer & 0 - 2 			\\ % & n \\
                      1 & on-site	&GER & Electronics 		& Architect	& \textgreater 10 	\\ % & n \\	
                      2 & on-site	&GER & Electronics 		& Developer & 3 - 5 			\\ % & y \\
                      2 & on-site	&GER & Electronics 		& Developer & \textgreater 10 	\\ % & y \\
                      2 & on-site &GER & Electronics 		& Developer & 0 - 2 			\\ % & y \\
                      2 & on-site &GER & Electronics 		& Developer & 6 - 10 			\\ % & y \\	
                      3 & on-site &GER & Manufacturing 	& Developer & \textgreater 10 	\\ % & y \\
                      3 & on-site &GER & Manufacturing 	& PM/PO		& \textgreater 10 	\\ % & y \\
                      3 & on-site &GER & Manufacturing 	& Developer & \textgreater 10 	\\ % & y \\
                      3 & on-site &GER & Manufacturing 	& Developer & 0 - 2 			\\ % & y \\
                      4 & on-site &GER & Insurance		& Developer & 0 - 2 			\\ % & y	\\
                      4 & on-site &GER & Insurance		& Developer & 3 - 5 			\\ % & y \\
                      4 & on-site &GER & Finance			& Developer & 0 - 2 			\\ % & y \\
                      4 & on-site &GER & Insurance		& PM/PO		& \textgreater 10 	\\ % & y \\
                      5 & online &GER & Transportation 	& IT-Manager& \textgreater 10 	\\ % & y \\
                      5 & online &GER & IT Consulting	    & Business	& 3 - 5 			\\ % & Y \\
                      5 & online &GER & IT Consulting	    & Architect	& \textgreater 10 	\\ % & y \\
                      5 & online &GER & Telecommunic.     & IT-Manager& \textgreater 10 	\\ % & y \\
                      6 & online &GER & Finance			& PM/PO		& 3 - 5 			\\ % & y \\
                      6 & online &GER & IT Consulting	    & Developer & 3 - 5 			\\ % & y \\
                      6 & online &GER & Finance			& Developer & 3 - 5 			\\ % & y \\
                      6 & online &GER & Finance			& Developer & 0 - 2 			\\ % & y \\
                      7 & online &GER & Energy        	& PM/PO    	& \textgreater 10   \\ % & y \\
                      7 & online &GER & E-commerce        & Developer & 0 - 2             \\ % & n \\ 
                      7 & online &GER & E-commerce        & Developer & 3 - 5 	        \\ % & n \\ 
                      7 & online &GER & Pharmacology      & IT-Manager& \textgreater 10   \\ % & y \\
					 8 & online &BRZ & E-commerce		 & Developer & 0-2			     \\ % & y \\
					 8 & online &BRZ & E-commerce        & Developer & 0-2               \\ % & y \\
					 8 & online &BRZ & Academic	   	     & Architect & 0-2               \\ % & y \\
					 8 & online &BRZ & Government	     & Developer & 0-2               \\ % & y \\
					 9 & online &BRZ & IT Consulting	 & PM/PO     & \textgreater 10   \\ % & y \\
					 9 & online &BRZ & E-commerce	     & Developer & \textgreater 10   \\ % & y \\
					 10 & online &BRZ & Finance	         & Developer &  6 - 10           \\ % & y \\
					 10 & online &BRZ & E-commerce	     & Developer & 0-2               \\ % & y \\
					 10 & online &BRZ & E-commerce	     & Developer & 0-2               \\ % & y \\
					 10 & online &BRZ & Finance	         & IT-Manager& \textgreater 10   \\ % & y \\
                      11 & online & BRZ & Finance			& PM/PO	 	& \textgreater 10 	\\ % & y \\
                      11 & online & BRZ & Finance			& User		& \textgreater 10	\\ % & y \\
                      12 & online & BRZ & Finance			& PM/PO		& \textgreater 10	\\ % & y \\
                      12 & online & BRZ & Finance			& PM/PO		& \textgreater 10	\\ % & y \\
                      12 & online & BRZ & Finance			& PM/PO		& \textgreater 10	\\ % & y \\
                      13 & online & PL & Ecommerce		& Architect	& \textgreater 10	\\ % & y \\
                      13 & online & PL & Finance			& IT-Manager& \textgreater 10	\\ % & y \\
                      13 & online & PL & Electronics   	& Developer	& 6 -10	            \\ % & y \\

                    \hline    		
                \end{tabular}
                }
    	\end{table}
        
        \subsection{Sample}
        \label{sec:sampling_strategy}
        A total of 46 practitioners played the game in 13 game sessions, of which nine were played online.
        Four games were played on-site (using the PDF version), by players who work together as a team or collaborate within their company, aligning with the study's initial goal to foster discussion on embedding TD management processes in a team. 
        For the nine online games, we used our personal network and assembled interested players promoting the game on social networks.
        Seven games were played in Germany, five in Brazil, and one in Poland. 
        In two of the games, two participants played as a team against a researcher, and in two other games, a researcher played alongside one practitioner against other practitioners. 
        We included these games in the study since they represent sessions that may occur in real life, e.g. some players may prefer to play with an outsider experienced with the game, or they may struggle to find exactly four players.
         %We assume that the setting is similar to the other games as the participants were previously unfamiliar with the game, and the researchers were not forcing a particular game strategy. 
        
        %CUT candidate:
        Ultimately, the players comprised one customer, one user, nine project or product managers, six IT managers, four software architects, and 25 developers. 
        Only four participants had not heard of the term TD before the game.
        An overview of all players is given in~\cref{tab:participants}.
        
        To compare the effect the game had on different stakeholder types, we categorized the players into three stakeholder categories:
        \begin{itemize}
            \item \textit{Business stakeholders:} Business management or customers, users, and project managers or product owners as their deputies in software development projects.
            \item \textit{Junior technical stakeholders:} Developers, architects, and IT managers with up to 5 years of experience.
            \item \textit{Senior technical stakeholders:} Developers, architects, and IT managers with over 5 years of experience.
        \end{itemize}
        Of the 35 technical stakeholders, 17 were senior stakeholders with more than 5 years of experience.

        \subsection{Data collection}
        \label{sec:data_collection}
        Each study's session comprised four steps:
        \begin{itemize}
            \item a 15-minute long introduction to the game and its rules.
            \item a 15-minute test game to understand the rules and develop a strategy
            \item 60 minutes or until one team has completed all three modules to play the (main) game
            \item 15 minutes to fill out the questionnaire
        \end{itemize}
        For two games, the time for the main game had to be reduced to 30 minutes for organizational reasons.
        We recorded each main game session and transcribed the recordings to allow us to explore intriguing discussions. 
        
        Additionally, each participant answered a questionnaire after the game. 
        First, the questionnaire asked for demographic information on the participant, e.g., domain or experience.
        Second, we asked about each of the \textit{aha}-moments (see~\Cref{tab:ahamoments}) that guided the game design. %, whether the participant got this recognition from the game or not, and whether the participant knew about this fact beforehand or not. 
        For each  \textit{aha}-moment, the participants could choose one of the following options:
        \begin{itemize}
            \item ``I had the AHA moment, but the realization is NOT NEW to me.'' % (YNN).
            \item ``I had the AHA moment, and the realization is NEW for me.'' %(YN). 
            \item ``I didn't have the AHA moment, but the realization is NOT NEW for me.'' %(NNN).
            \item ``I didn't have the AHA moment, and the realization would be NEW for me.'' % (NN). 
        \end{itemize}
        Third, we asked whether the game changed the participant's attitudes and whether they plan to change their behavior regarding TD using two open questions.
        Finally, we asked for feedback on the game and ideas for its improvement.
              
        \subsection{Data Analysis}
        \label{sec:data_analysis}
        We quantitatively analyzed the questionnaire's data on the \textit{aha}-moments using Excel and created descriptive statistics for the results. 
        We do not expect all participants to experience all \textit{aha}-moments. 
        Depending on experience and background, different topics might be relevant to the players, and thus, other insights might be memorable to them.
        For example, while a business stakeholder might recognize or be surprised by the outer viscous cycles, a developer might be more surprised about the business causes and chains of causes.
        %CUT candidate:
        %Consequently, the use of inferential statistics, e.g., to identify whether a statistically significant amount of players had one \textit{aha}-moment is not reasonable.

        We qualitatively analyzed the answers to the open questions and the recordings' transcriptions. 
        To track changes in attitudes and behaviors, we counted responses that indicated a change or a plan to change. We also noted how many responses stated no intention to alter their current attitudes or behaviors.
        Furthermore, we identified repeated explanations for their changes in attitudes and behavior and also counted their frequency.
        For the feedback on the game, we counted the number of players who expressed their satisfaction with the game and identified further potential for improvement.
        As these three questions were optional, some players skipped them, leading to fewer responses than the total number of players.
        Finally, we read through the recordings' transcriptions and marked interesting discussions and remarks that showed participants could relate the game to their real-life experiences.
        
        To answer RQ 1 on the emulation of real-life experiences, we utilized the \textit{``I had the AHA moment, but the realization is NOT NEW to me.''} answers and combined those with the answers to the open questions and the transcriptions' analysis.
        To answer RQ 2 on the new insights on TD and TD management, we utilized the \textit{``I had the AHA moment, and the realization is NEW for me.''} answers and also combined those with the answers to the open questions.
        For RQ 3 on the changes in attitude or planned behavior, we utilized a qualitative analysis of the answers to the open questions.
    
    \section{RESULTS}
    \label{sec:Results}
    
    In this section, we present the results of our evaluation by answering each research question.
 %   We give a detailed overview of the \textit{aha} moments and present the relevant aspects of the qualitative analyses providing citations of discussions or the open questions. 

    \begin{figure*} 
        \centering
        \begin{tabular}{@{}c@{}}
            \subfigure[over all players]
            {	\label{fig:all} 
                \includegraphics[width=0.7\textwidth]{"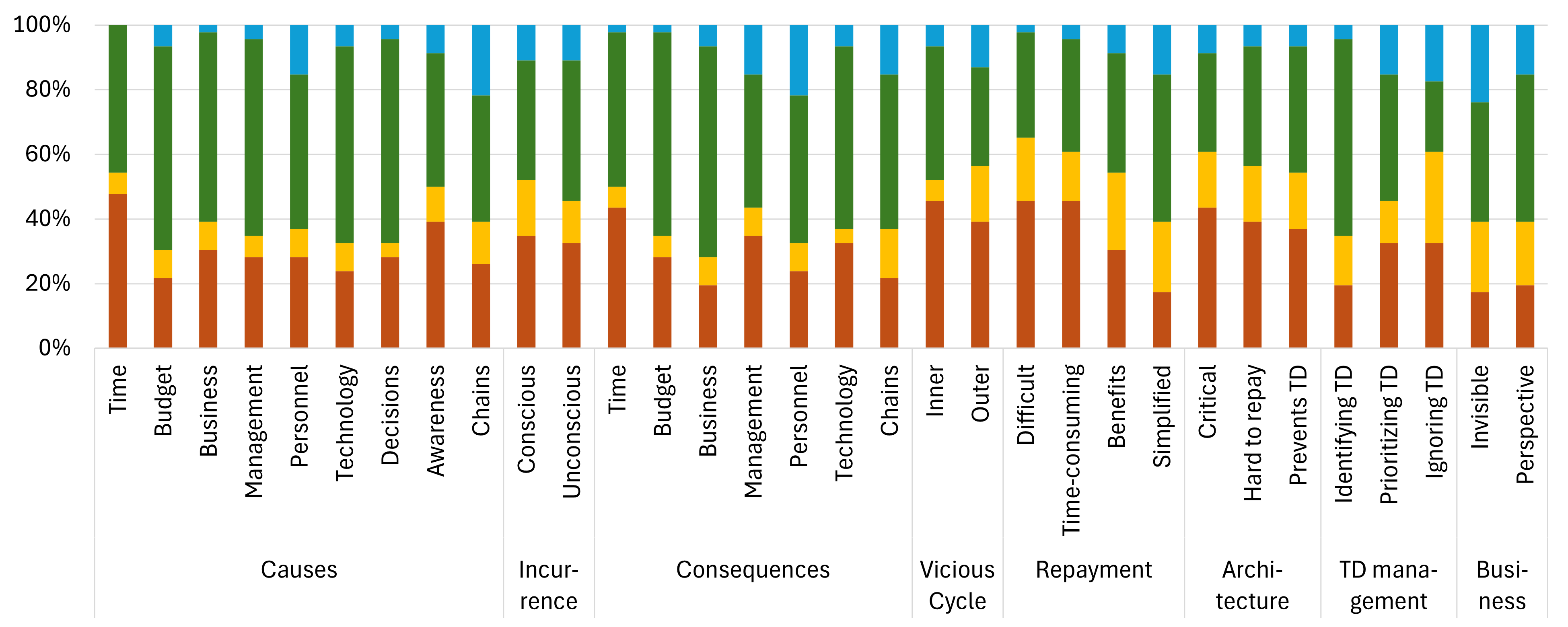"}
            }\\
            \subfigure[business stakeholders]
            {	\label{fig:business} 
                \includegraphics[width=0.7\textwidth]{"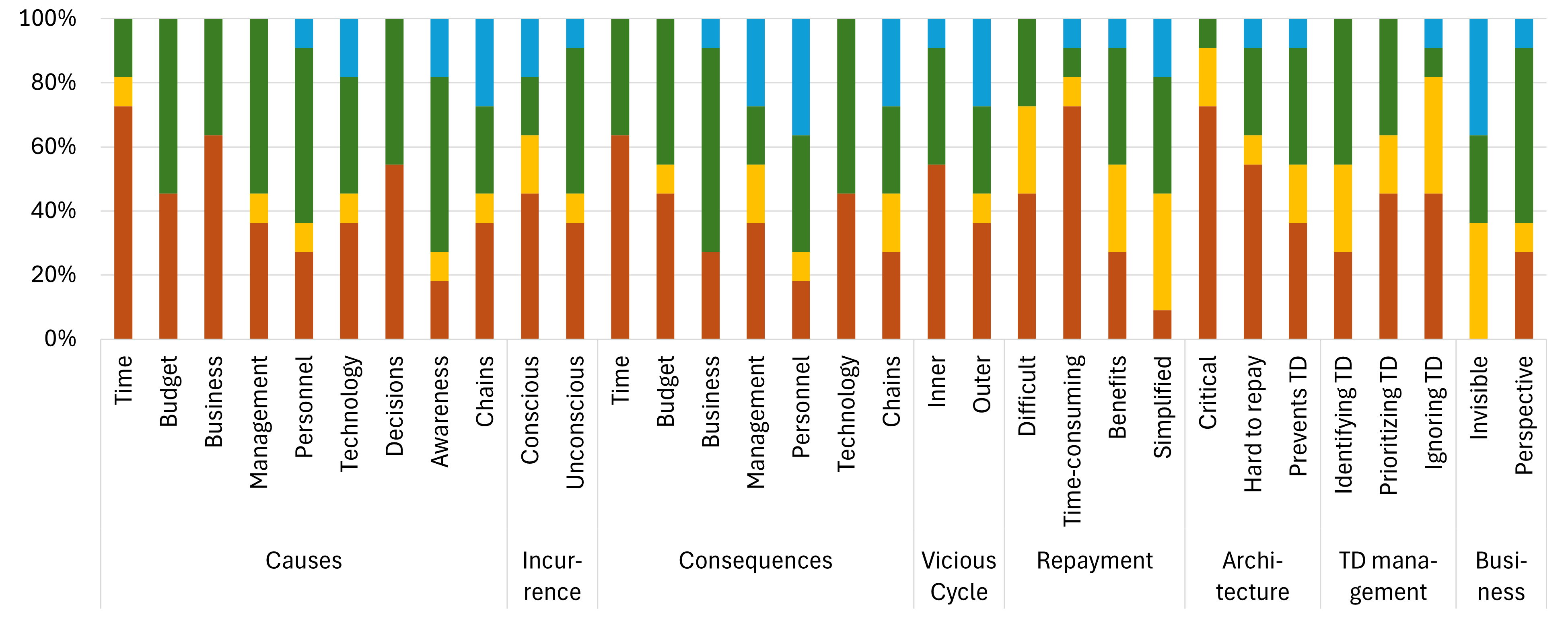"}
            }\\
            \subfigure[junior technical stakeholders]
            {	\label{fig:juniors} 
                \includegraphics[width=0.7\textwidth]{"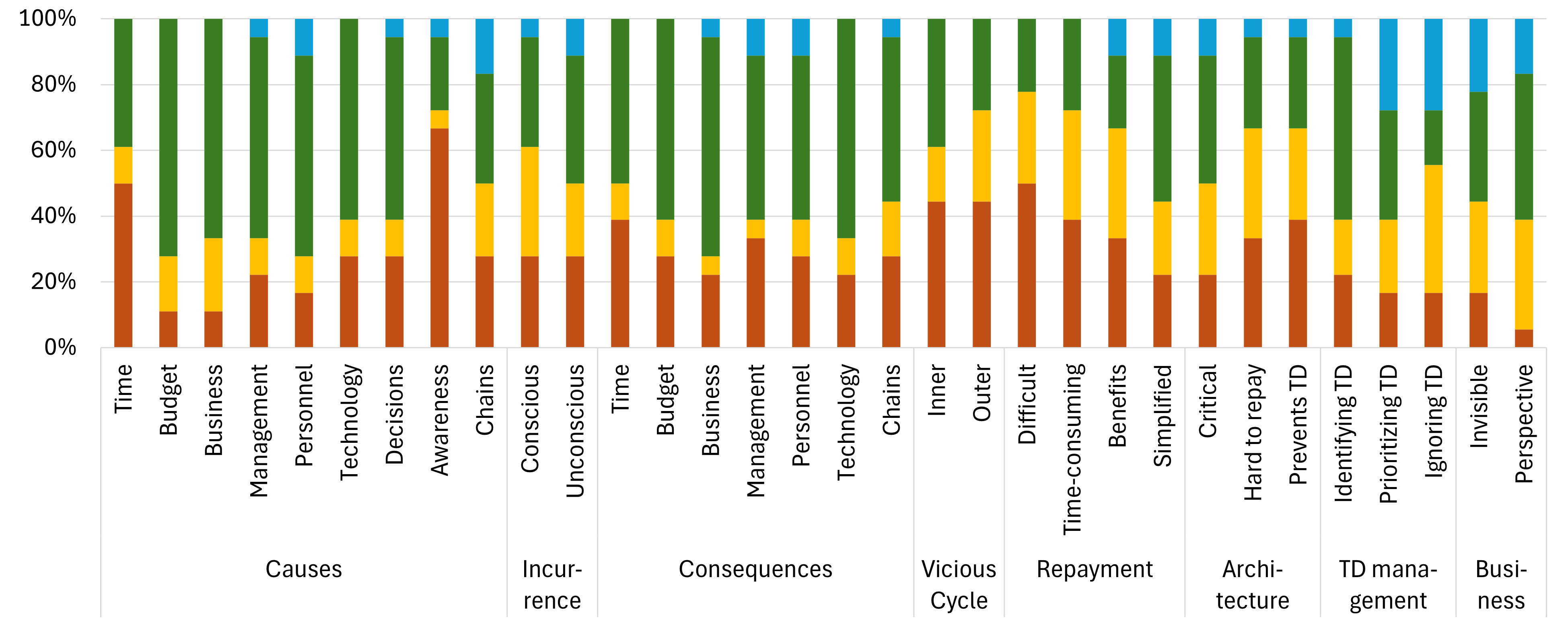"}
            }\\
            \subfigure[senior technical stakeholders]
            {	\label{fig:seniors} 
                \includegraphics[width=0.7\textwidth]{"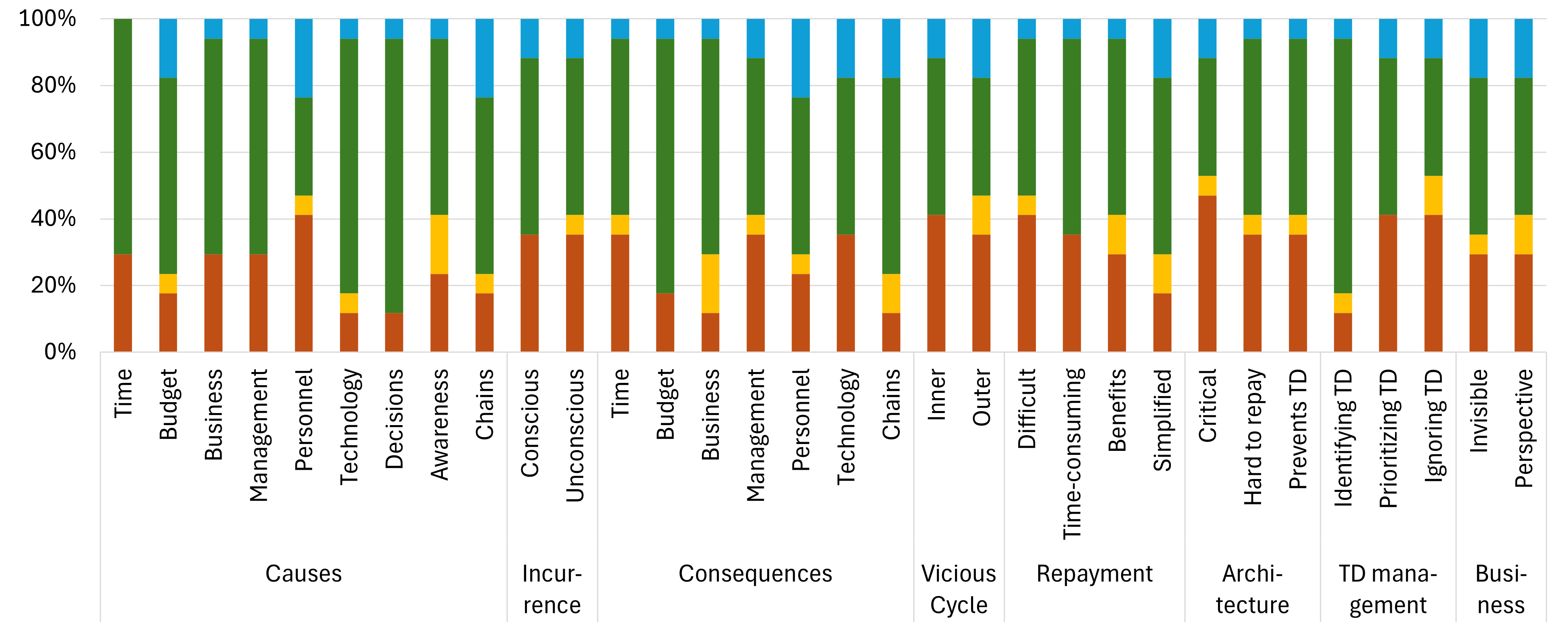"}
            }\\
            \subfigure[senior technical stakeholders]
            {	\label{fig:seniors} 
                \includegraphics[width=0.7\textwidth]{"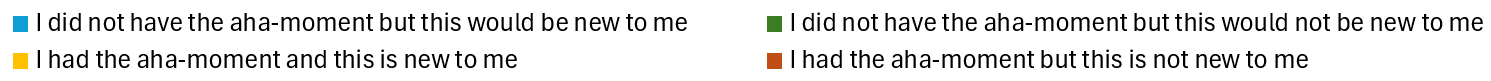"}
            }

        \end{tabular}

        \caption{Percentage of players that had the respective aha-moments for all players and per stakeholder category}
        
    \end{figure*}
            
    \begin{figure}%[H]
        \centering
        \includegraphics[width=0.4\textwidth]{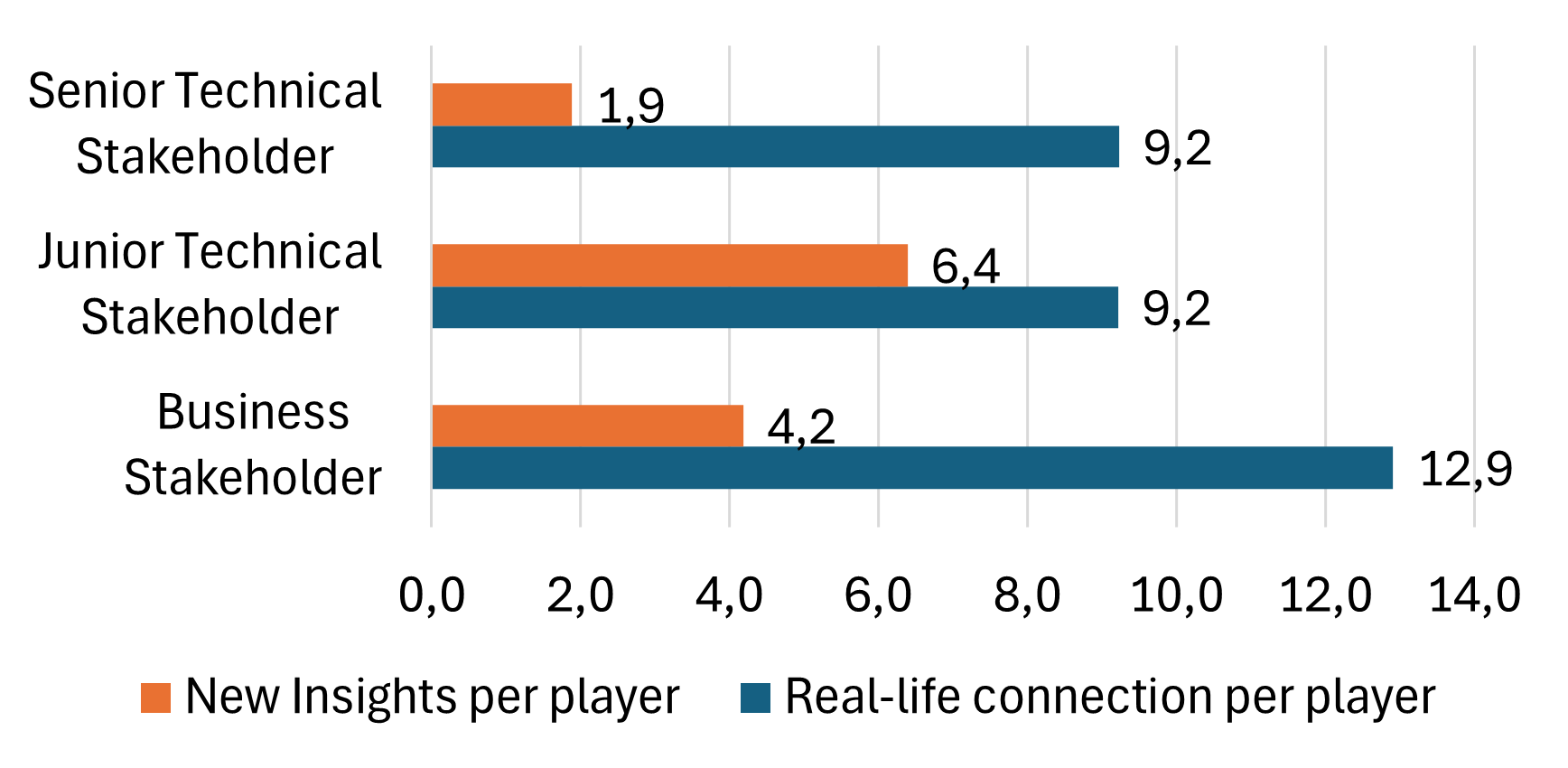}
        \caption{Average new insights (I had the aha moment, and it is new to me) and average real-life connections (I had the aha moment, and it is not new to me) per stakeholder category.}
        \label{fig:insights_reallife}
        %\vspace{-3mm}
    \end{figure}

        %----------------------RQ1 - Discussion ------------------------
        \subsection{RQ1: To what extent does the game emulate the real-life experiences that technical stakeholders encounter?}
        \label{sec:Results-RQ1}
        \Cref{fig:seniors} shows that for all of the \textit{aha}-moments, only a few senior technical stakeholders did not recognize them from their previous experience (yellow + blue bars), which demonstrates that the \textit{aha}-moments represent real-life situations.
        Furthermore, \Cref{fig:all} shows that for each aha-moment, 30\% to 60\% of all players had the respective \textit{aha}-moment (orange + yellow bars). 
        \textit{Aha}-moments for vicious cycles, repayment, architecture, and TD management were experienced most often (yellow + orange bars), while causes and consequences were recognized less frequently. 
        Particularly, the players rarely experienced the \textit{aha}-moments of budget causes and business consequences.
        These low values might be addressed by adding more action and event cards to the game. 
        As all \textit{aha}-moments were perceived by at least some players, we assume that the game offered the opportunity to experience all targeted \textit{aha}-moments.
        %Nonetheless,  the \textit{aha} moments that each stakeholder remembers after the game are subjective to their individual experiences.

        \Cref{fig:insights_reallife} also shows that from each stakeholder category, each player could relate at least nine \textit{aha}-moments to their real life, further proving the realistic design of the game.
        
        Further, we identified that connections to real life were made during the games, as the following examples show:
%        \begin{itemize}
%            \item 
            \textit{``%P2: Mhm, if you don't take on technical debt. /
            P1: We already have so much [TD]in the system. / P2: That's always one [TD per feature]. (Laughter) / P1: It's really like reality. (Laughter)''}  and % / P1: Should we incur another one? / P2: Okay, if we have five tasks to complete.''}
            \textit{``P1: [Card is read out] Developer retires. A good developer retires. (Laughs) He wrote good code but, unfortunately, didn't document it very well. Oh, that sounds familiar. (Laughs)''}
       % \end{itemize}
        
        Finally, the survey's feedback on the game included statements supporting the game's realism: % of  (questionnaire) of a senior technical stakeholder:
        \textit{``I liked how it [the game] simulates the difficulties and the setbacks we often find during software development.''} %CUT: The game does a good job of showing how TD impacts the development process and how it's important to, if not eliminate, at least keep it manageable.''}
        \textit{``The ?-cards were sometimes quite hard and set you back in the game, but that's how it is in reality with technical debt, so they are justified.''}
        Some more critical feedback involved the ease of repaying architectural TD:~\textit{``But I do not like that repaying in this game TD of architecture TD has 50\% of success [b]ecause I think repaying TD in architecture Level is much worse to do.''}
        This feedback can be used to adapt the game, e.g., by blocking more digits to repay architectural TD.

        Regarding RQ 1, the \textit{aha}-moments, the transcriptions, and the feedback led us to consider the game as a realistic emulation of the real world.

        \subsection{RQ2: Do the players encounter new insights and learn about additional facets of TD management?}
        \label{sec:Results-RQ2}

        \Cref{fig:all} shows the \textit{aha}-moments that were new to the players in yellow. 
        All \textit{aha}-moments were new to at least one player showing the relevance of each \textit{aha}-moment.
        Notably, there is a substantial difference between stakeholders, which is also revealed in~\Cref{fig:insights_reallife}. 
        While senior technical stakeholders encounter, on average, less than two new insights, i.e., new \textit{aha}-moments, business stakeholders more than four, and junior technical stakeholders encounter more than six.
        This illustrates that the game might assist in harmonizing the level of knowledge of all stakeholders.
        % Moved to discussion:
        %Academic literature shows that TD incurred by inexperienced developers is one frequently mentioned cause of TD~\cite{Freire2021a, wiese_it_2023}. 
        %It also reveals that communication with business stakeholders can be challenging~\cite{Ernst2015, Besker2018b, wiese_it_2023}.
        %Unnecessary repetition:
        %Thus, we assume that the game supporting their understanding of TD is a noteworthy contribution.

        Business stakeholders often noted new insights regarding measures that simplify refactorings, ignoring TD as a possibility and the fact that some TDs are invisible to business stakeholders and can only be noticed by their symptoms (yellow bars in~\Cref{fig:business}).
        Further, the difficulties in repaying TD and identifying it were new to multiple business stakeholders.
        This indicates that crucial aspects of their work in determining which TD items should be prioritized as high or low in the backlog may have been previously overlooked. 
        %CUT candidate:
        %Nonetheless, due to the low amount of business participants, the data has to be evaluated carefully and may only indicate the potential effects the game might have on this stakeholder category.
        %CUT - part of next question:
        %Furthermore, most business stakeholders indicated that their attitude towards TD remained unchanged because they had frequently encountered the topic of TD and were managing it to the best of their abilities.
        %Only one business stakeholder stated that they were \textit{``reminded of how best to treat TD. And how hard it is to get rid of them''} and they plan to keep a better overview in the future.
        % Only one business stakeholder mentioned the highlighted \textit{aha}-moments in their feedback: \textit{``Although some TD is invisible, there are enough visible TD (clear decision made for this) or those that become visible, which should also be clearly named.''}

        For junior technical stakeholders, the new insights are broadly distributed across all \textit{aha}-moments  (yellow bars in~\Cref{fig:juniors}).
        The topics of TD incurrence, vicious cycles, TD repayment, the relevance of TD in architecture, TD management, and the business perspective on TD are often perceived as new insights by them. 
        At the same time, they less frequently recognized \textit{aha}-moments on causes and consequences.
        This indicates that the broad perspective on TD embodied in the game is a valuable way of making TD knowledge tangible for junior technical stakeholders.
        The effect on junior stakeholders is also mirrored in their answers to the open questions: \textit{``If you have no experience with the topic of TD, the game could very well help you learn the basics in a playful way''} 
        %CUT candidate:
        %and \textit{``Every piece of information is valuable, especially for newcomers to the IT field.''} %CUT: . . the game is very interesting and certainly helps to give a good insight into common problems in software development.''}
        %CUT: A more detailed summary is given by another junior stakeholder: \textit{``A fixed time limit for a game run clearly conveys the impact that time pressure can have and how TD can arise as a result. The dependencies clearly show that old TD causes new ones or that new developments take longer. It was also very good that it became clear that time was needed to eliminate TD. ''}

        For RQ 2, we summarize that while all players gain new insights, business stakeholders and junior stakeholders acquire the most knowledge from the game.
        In the end, this harmonizes the knowledge about TD concepts between stakeholders, facilitating knowledgeable discussions on TD management after the game.

        In the end, this harmonizes the knowledge about TD concepts between stakeholders, enabling knowledgeable discussions on TD management after the game.
    
        \subsection{RQ3: How does the game change players' attitudes and possibly behaviors regarding TD?}
        \label{sec:Results-RQ3}
    
        Out of all 46 players, 26 of 44 responding players replied that they changed their attitude towards TD, and 27 of 34 responding players plan to change their behavior.
        
        %attitude change
        Change in attitude comprises (number of mentions in brackets) more awareness of causes (5) and consequences (11), repayment of TD (7), dependencies between TD items (5), the importance of TD management (2), easier TD identification (2) and the significance of architectural TD (1).
        For example, one player said: \textit{``The game clearly shows what consequences TD can cause in a module and how important it can be to fix TD. That's why the game sensitized me.''}
        Another mentioned: \textit{``It made me realize that the impact of TD on subsequent tickets is not negligible and should be dealt with strategically.''} %CUT: It can limit the scope for action more and more.''}
        Additionally, five players recalled the benefits of incurring TD for swift product delivery.
        One of them mentioned that TD is \textit{``[n]o longer a concept of fear because it was conveyed that you can deal with it or work with it.''}
        
        Among the 18 players who did not change their attitude, 17 reported that they were already familiar with the TD concepts and how they work, therefore, they felt no need to alter their attitude.
        One technical stakeholder stated: \textit{``My attitude hasn't changed; technical debt is part of the core business of my job.''}
        A business stakeholder noted: \textit{``. . . I had to fight for many years against management that prioritized time for short-term features above all TD.''} % . . . %Thanks to a stable architecture, we could compensate for a lot in the project, but 
        % today, I would name and track the TD more clearly to increase transparency.''}

        %behavior change
        Planned behavior changes (number of mentions in brackets) comprise making decisions on TD incurrence more rational by evaluating options and their consequences (12), employing a more structured approach to TD (10), starting tracking TD (7), repaying TD more regularly (3), and being more aware of TD overall (4).
        For example, one player said, regarding TD incurrence, that they \textit{``plan to look for TDs in a more critical way and see how it can impact the development process in the future.''}
        Regarding tracking, one player noted that \textit{``[a]lthough some TD are invisible, there are enough visible TD . . . %(clear decision made for this) 
        or those that become visible, which should also be clearly named.''}
        Concerning a structured TD management approach, one player noted \textit{``. . . it seems even more important to me to manage technical debt thoughtfully.''}

        Two players who did not plan to change their behavior mentioned other people or institutions, like  the management or the company, that already manage TD, e.g.,~\textit{``. . .  in my context, there are already processes for dealing with technical debts and . . . they are under the responsibility of other people.''}
        Five players noted they did not plan to change their behavior but stated they were already aware of TD and they %tried to avoid and repay them whenever possible: \textit{``I  
        \textit{``always try to avoid technical debt and only give in when pressure comes from above.''}

        For RQ 3, we conclude that most players are open to changing their attitude and behavior regarding TD and TD management after playing the game.
        %Further, nearly all of the players who were not open to changing their attitudes or behaviors reason that this is because they already know about and manage TD. 
        Furthermore, almost all players unwilling to change their attitude or behavior justified this by claiming they already knew about TD and tried to manage it. 
        In both cases, the players were open to productive and rational discussions, which was one of the game's goals.
            
    \section{DISCUSSION}
    \label{sec:Discussion}
    In this section, we discuss the results of ~\Cref{sec:Results} in terms of the game's effect on different stakeholders. 
    Additionally, we summarize the feedback on the game itself.
    
        \subsection{Effect on Senior Technical Stakeholders} % filled the gaps
        Senior technical stakeholders got the least new insights from the game. However, we were still able to fill some knowledge gaps, e.g., for three seniors, the fact that TD might have consequences for the business was new, which is important new knowledge.
        Senior technical stakeholders also exhibited the least change in attitude towards TD since they already had vast TD knowledge. 
        Nonetheless, many of them were open to behavior changes after playing the game, and ideas on what to change were reinforced. %, which was the basic goal of the game.
        Besides filling knowledge gaps, the most significant benefit of having senior technical stakeholders participate in the game is sharing their knowledge with junior and business stakeholders. 
        This exchange might be a valuable starting point for TD management discussion.
        % Likewise, the effect of being able to share their experience with junior and business stakeholders during the game is noteworthy. 
        % Not only can they transfer their knowledge, but the exchange might also function as a starting point for further discussions on managing TD.
        
        \subsection{Effect on Junior Technical Stakeholders} % not everything is perfect (ignoring is an option)
        Of all stakeholders, junior technical stakeholders profited the most from new insights.
        The fact that sometimes ignoring TD is a valid option was new to twelve of them, and seven mentioned that they had this \textit{aha}-moment. 
        Being still new to IT, developers might be looking for the perfect solution and lack the experience to sometimes accept compromises.
        The game taught them that compromises must sometimes be made to reach business goals.
        %CUT: In the open questions, one junior developer summarized their learning: \textit{``A fixed time limit for a game session clearly conveys the impact that time pressure can have and how TD can arise as a result. The dependencies clearly show that old TD items cause new ones or that new developments take longer. It was also very good that it became clear that time was needed to repay TD.''} %CUT: The influence of resources on the TD was more difficult because there were no resources (money, personnel) to manage.''}
        Additionally, academic literature shows that TD incurred inadvertently by inexperienced developers is one frequently mentioned cause of TD~\cite{Freire2021a, wiese_it_2023}. 
        The game might support junior developers in being more careful during implementations and showing more understanding in discussions with experienced developers.
        
        \subsection{Effect on Business Stakeholders} % They do not see everything and have to trust technical people
        Academic literature reveals that communication with business stakeholders can be challenging~\cite{Ernst2015, Besker2018b, wiese_it_2023}.
        By playing the game, business stakeholders had substantially more new insights than senior technical stakeholders and thus learned new characteristics of TD, which might be beneficial in future discussions about TD prioritization.
        In particular, the reality that some TD might not be visible to them was new to eight of eleven business stakeholders and mentioned by four of them as a new learning. 
        Providing this crucial knowledge might be useful for business stakeholders to put more trust in the technical stakeholders' evaluation of TD.
        This may be a particularly useful effect of the game since missing business stakeholders' trust makes communication about TD especially difficult~\cite{wiese_it_2023}.
    %     As missing trust by business stakeholders was one outcome of recent research~\cite {wiese_it_2023}.

    % Missing trust by business stakeholders was one outcome of recent research~\cite {wiese_it_2023}.
    
        % However, three business stakeholders stated they were already aware of the TD concepts before the game.
        % Therefore, only eight of the eleven business stakeholders changed their attitude toward TD, and only six planned to modify their behavior after the game.  
        
        % Additionally, it is distinctive that two of the eleven business stakeholders, compared to two of the technical stakeholders, raised concerns about too much luck influencing the game, suggesting that business stakeholders may be more competitive than other participants.
        % In that case, losing the game might lead to frustration, hindering discussion about TD, which could be analyzed more thoroughly in future game sessions.
        
        %Regardless, we cannot make any conclusive assumptions about the game's impact on business stakeholders due to the low number of business stakeholders and the indecisive data.

        \subsection{Feedback on the TechDebt Game} 
        The overall feedback on the game was encouragingly positive.
        %fun
        15 players explicitly stated that they had fun playing the game, e.g., \textit{``The game is a lot of fun! I would also play privately!''} %CUT: , \textit{``I would like to purchase the printed version.''}  
        or \textit{``With just a few minutes playing the game, I felt immersed and could think about some strategies.''} % to gain more points.''}
        %CUT: More specifically, one player wrote: \textit{``What I find exciting about the game is that it was fun and still provides the above-mentioned insights [\textit{aha}-moments].''} %CUT: or makes them more aware.''}
        Even while playing the game, some players wanted to immediately ``buy'' the game:
        \textit{``P2: . . . I think it's a really good idea as a board game. . . . %I think you could make a really good board game out of it. / P1: Yes. / P2: Even with multiple teams and so on. 
        I would buy it. (Laughs) / P1: Ah, okay. New Ravensburger [German  game manufacturer] Technical Debt game.``}
        Accordingly, after one of the games, we even got an offer to support professionalizing the game.
        Moreover, one company was considering using the game company-wide to change the company's culture regarding TD.

        %Improvement
        Some players made suggestions for improvement, e.g., regarding TD causes, one player mentioned that
        \textit{``%Of course, the game has limitations in showing the consequences for business and management. 
        [t]he game was able to convey the “time” factor for the creation of TD very well, but the “resources” factor (money, personnel) less well. The event cards were good for this, to get examples.''}
        Regarding the action and event cards, two players mentioned that \textit{``%it was a lot of fun and can be surprisingly diverse when it comes to strategies. 
        [m]ore different '!' [action] or '?' [events] effects would make it [the game] more interesting.''}
        %CUT: \textit{``I think the cards with a lightning icon could have more different effects. They have a unique text description, but the effects are always the same two or three.''}
        Another player mentioned: %CUT: expects that more cards would have led to further \textit{aha}-moments:
        \textit{``The event cards were a nice way to impart knowledge about TD, and I believe many of my `I did not have the aha moment' would have appeared through other cards in the game.''}
        % Few players mentioned that repaying architectural TD was to easy:~\textit{``But I do not like that repaying in this game TD of architecture TD has 50\% of success [b]ecause I think repaying TD in architecture Level is much worse to do.''}
        One player missed interaction: \textit{``The only thing I missed was more interaction with the other teams in the game.''} 
        Finally, as mentioned in the previous sections, some players also complained about the influence of luck and the ease of repaying architectural TD. 
        %CUT: Sixth, one player mentioned that they would have preferred a \textit{``tracking based on moves (sprints) instead of time''}. 
        %However, we specifically removed this part from version 1 due to other players complaining about the complexity of this element.
        %CUT:
        %Finally, one player stated that \textit{``[h]ere and there, the balancing in the game doesn't match my practical experience (reducing TD in the architecture is very expensive in my projects), but that doesn't have to be wrong.''}
        %CUT: \textit{``definition of time for certain moves, ways to separate the time to allow each team to play.''}
        
        %While the feedback uncovers some ideas for improvement, the game overall was fun to play

    \section{THREATS TO VALIDITY}
    \label{sec:ThreatsToValidity}
    	We present threats to validity based on the guidelines provided by Wohlin et al.~\cite{Wohlin2012}.
    	
    	\paragraph{\textbf{Construct Validity}}
            % Construct validity refers to the appropriateness of the chosen methods.	
            % * When we design the study
            % * potential problems of the design of our study
            First, we built the game along the \textit{aha}-moments as learning objectives. 
            We derived these moments from scientific literature, e.g., the works of Li et al.~\cite{Li2015}, Wiese et al.~\cite{wiese_it_2023}, Martini et al.~\cite{Martini2015a}, Kruchten et al.~\cite{Kruchten2019}, and Ramac et al.~\cite{ramavc2022prevalence}.
            While we might have missed important aspects of TD and TD management, the game is designed in a way that might embed further \textit{aha}-moment by creating additional action or event cards.
            Second, in four game sessions, we deviated from the original study design and let a researcher play against or alongside a participant, % as it was sometimes difficult to organize appointments with four participants, which posed 
            posing a threat to construct validity.
            However, this may improve external validity as these game sessions replicate real-life situations.
            %We consider the setting similar enough to the other game sessions. 
            %, as the researchers were not forcing a particular game strategy.

    	\paragraph{\textbf{Internal Validity}}
            % Internal validity threats are related to possible wrong conclusions about causal relationships between treatment and outcome. 
            % Threats in the process of the studies execution
            First, to analyze the game's effectiveness, we analyzed the subjective perspective of the participants and did not provide objective measures. 
            We assume that this is in line with the overall goal of the game, which is to serve as a starting point for discussing TD and to foster an open yet educated mindset on this topic, which is a highly subjective perception.
            Second, the qualitative analysis of transcripts and open-ended questions may introduce biases. 
            To mitigate this, two researchers collaboratively reviewed all results. 
            Additionally, the answers to the open questions and their analysis are part of the additional material~\cite{AdditionalMaterial} to be reviewed by all readers of this paper.
            We also combined qualitative findings with quantitative survey data to address the research questions.
            Finally, one game goal is to foster the discussion about TD, which we did not analyze specifically. 
            We derived the openness for discussion by analyzing participants' willingness to change their behavior and added some citations proving that discussions indeed happened. 
            However, this is subjective or anecdotal knowledge and may be evaluated further by analyzing team discussion after playing the game.
    	
    	\paragraph{\textbf{External Validity}}
    	% External validity threats refer to the ability to generalize the result.
            % how generalizable our findings are
            First, we played the game with participants from more than ten different industry domains and included various stakeholders of many experience levels.
            However, like most industry research, our sample is accompanied by selection bias, as mainly players interested in the TD topic were willing to participate.
            Second, we would have preferred to include more business participants in the evaluation.
            Yet, in real life, one project manager or product owner manages multiple developers, which is a ratio that is also exhibited in our game sessions.
            Third, none of the games was played without a knowledgeable game master. 
            Thus, we cannot ensure that the game instructions document is sufficiently self-explanatory. 
            However, the players used the manual to examine rules, and we optimized the manual throughout the various iterations, resulting in an additional short manual and an overview on the game board.
    
    	\paragraph{\textbf{Conclusion Validity}}
    	% Conclusion Validity relates to the reliability of the conclusions drawn from the results.    
            % * after we have executed the study
    	% * limitations of our data analysis
            Our conclusions are only short-term related, i.e., we only surveyed the player's situation right after they played the game. 
            This is aligned with our goal of starting discussions on TD management. 
            We did not systematically analyze whether long-term changes resulted from the change.
            %CUT: We know that three of the teams that played the game embedded TD management into their processes afterward.
            % However, this was guided by another research project with multiple interventions, and thus, the long-term effect of the game cannot be deduced from this knowledge. 

\section{CONCLUSION}
\label{sec:Conclusion}

%Todo
        In this paper, we present a board game for conveying TD concepts, making experiences with TD tangible, and fostering discussions on managing TD. % along the software development life cycle.
        The game was designed based on a thorough theoretical foundation presented in~\Cref{sec:background}. 
        We evaluated the game by playing it with 46 practitioners in 13 game sessions using a questionnaire after the game to gather feedback on the conveyed \textit{aha}-moments.
        For qualitative feedback, we added open questions to the questionnaire and analyzed transcripts from the games' recordings.

        By answering our RQs, we established that the game is fun to play, realistic, and can be used to harmonize the level of knowledge of all stakeholders.
        Many players changed their attitude toward TD and plan to change their behavior.
        %CUT candidate:
        %Behavior changes focus on making decisions on TD incurrence more rational, employing a more structured approach to TD, starting tracking TD, repaying TD more regularly, and being more aware of TD overall.
    
        \textbf{Practitioners,} can use the game to experiment with various strategies in a safe environment (``graceful failure'').  
        The game provides an unbiased and rational basis for further TD discussions. % ground on further TD management strategies.
        For long-term effectiveness, however, the game must be followed by a discussion on embedding TD management in the current software development life cycle.
 
        \textbf{For researchers,} we presented that game-based learning is not only a method to impart knowledge but also a means to foster discussions and spark behavior changes.
        The game might also be used as a starting point when training practitioners on methods to manage TD.

    %\subsection{Future Work}
    \textbf{Future work} might include researching how educational programs and training might help embed TD management in an existing software development life cycle. 
    The game might then be used as a first step in those trainings.
    Moreover, the company of one game session's participants considers using the game company-wide to foster a change of mindset throughout the company, which could be evaluated.
    In our study, players only played the game once. 
    A further research topic might be to evaluate how experimenting in this emulated environment might be profitable. 
   % Finally, the game can be improved by adding new \textit{aha}-moments.

%\section*{Acknowledgment and Funding}
\textbf{Acknowledgment and Funding: }
    We thank the game players for their participation, their trust, and the insights they provided.
    % and the students who were participating in the game development but did not want to co-author this paper.
    In memory of Prof. Dr.-Ing. André van Hoorn and Prof. Dr.-Ing. Matthias Riebisch.    The project on which this report is based was sponsored by the Federal Ministry of Education and Research of Germany under the funding code 01IS24031. Part of this work was supported by Funetec-PB – Ct.: Phoebus 01/24.
    Responsibility for the content of this publication lies with the authors.
%CUT candidate:
    % While preparing this work, the authors used Grammarly~\cite{Grammarly} to improve the readability and language of the manu\-script, f4x~\cite{f4x} to transcribe the recordings, and DeepL~\cite{DeepL} to translate German and Portuguese to English.
    % After using these tools, the authors reviewed and edited the content as needed. 
    % They take full responsibility for the content of the published article.

\bibliographystyle{IEEEtran}
\bibliography{TDGame}

% Generated by IEEEtran.bst, version: 1.14 (2015/08/26)
\begin{thebibliography}{10}
\providecommand{\url}[1]{#1}
\csname url@samestyle\endcsname
\providecommand{\newblock}{\relax}
\providecommand{\bibinfo}[2]{#2}
\providecommand{\BIBentrySTDinterwordspacing}{\spaceskip=0pt\relax}
\providecommand{\BIBentryALTinterwordstretchfactor}{4}
\providecommand{\BIBentryALTinterwordspacing}{\spaceskip=\fontdimen2\font plus
\BIBentryALTinterwordstretchfactor\fontdimen3\font minus
  \fontdimen4\font\relax}
\providecommand{\BIBforeignlanguage}[2]{{%
\expandafter\ifx\csname l@#1\endcsname\relax
\typeout{** WARNING: IEEEtran.bst: No hyphenation pattern has been}%
\typeout{** loaded for the language `#1'. Using the pattern for}%
\typeout{** the default language instead.}%
\else
\language=\csname l@#1\endcsname
\fi
#2}}
\providecommand{\BIBdecl}{\relax}
\BIBdecl

\bibitem{Avgeriou2016a}
P.~Avgeriou, P.~Kruchten, I.~Ozkaya, and C.~Seaman, ``{Managing Technical Debt
  in Software Engineering},'' \emph{Dagstuhl Reports}, vol.~6, no.~4, pp.
  110--138, 2016.

\bibitem{Kruchten2019}
P.~Kruchten, R.~Nord, and I.~Ozkaya, \emph{{Managing Technical Debt: Reducing
  Friction in Software Development}}.\hskip 1em plus 0.5em minus 0.4em\relax
  Software Engineering Institute, Carnegie Mellon University, 2019.

\bibitem{Ernst2021}
N.~Ernst, R.~Kazman, and J.~Delange, \emph{Technical {Debt} in
  {Practice}}.\hskip 1em plus 0.5em minus 0.4em\relax London, England: The MIT
  Press Cambridge, Massachusetts, 2021.

\bibitem{Li2015}
Z.~Li, P.~Avgeriou, and P.~Liang, ``{A systematic mapping study on technical
  debt and its management},'' \emph{Journal of Systems and Software}, vol. 101,
  pp. 193--220, 2015.

\bibitem{rios2018tertiary}
N.~Rios, M.~G. de~Mendon{\c{c}}a~Neto, and R.~O. Sp{\'\i}nola, ``A tertiary
  study on technical debt: Types, management strategies, research trends, and
  base information for practitioners,'' \emph{Information and Software
  Technology}, vol. 102, pp. 117--145, 2018.

\bibitem{Junior2022}
H.~J. Junior and G.~H. Travassos, ``{Consolidating a common perspective on
  Technical Debt and its Management through a Tertiary Study},''
  \emph{Information and Software Technology}, vol. 149, no. October 2020, p.
  106964, 2022.

\bibitem{wiese_it_2023}
M.~Wiese and K.~Borowa, ``\BIBforeignlanguage{en}{{IT} managers’ perspective
  on {Technical} {Debt} {Management}},'' \emph{\BIBforeignlanguage{en}{Journal
  of Systems and Software}}, vol. 202, p. 111700, Apr. 2023.

\bibitem{InsighTD2022}
\BIBentryALTinterwordspacing
``{InsighTD Project – InsighTD Project}.'' [Online]. Available:
  \url{http://www.td-survey.com/}
\BIBentrySTDinterwordspacing

\bibitem{Rios2020}
N.~Rios, R.~O. Sp{\'{i}}nola, M.~Mendon{\c{c}}a, and C.~Seaman, ``{The
  practitioners' point of view on the concept of technical debt and its causes
  and consequences: a design for a global family of industrial surveys and its
  first results from Brazil},'' \emph{Empirical Software Engineering}, vol.~25,
  no.~5, pp. 3216--3287, 2020.

\bibitem{Ramac2021}
R.~Ramač, V.~Mandić, N.~Taušan, N.~Rios, S.~Freire, B.~Pérez,
  C.~Castellanos, D.~Correal, A.~Pacheco, G.~Lopez, C.~Izurieta, C.~Seaman, and
  R.~Spinola, ``Prevalence, {Common} {Causes} and {Effects} of {Technical}
  {Debt}: {Results} from a {Family} of {Surveys} with the {IT} {Industry},''
  \emph{Journal of Systems and Software}, 2021.

\bibitem{Cunningham1992}
W.~Cunningham, ``{The WyCash portfolio management system},'' in
  \emph{Proceedings of the Conference on Object-Oriented Programming Systems,
  Languages, and Applications, OOPSLA}, vol. Part F1296, no.~2, 1992, pp.
  29--30.

\bibitem{Verdecchia2021}
R.~Verdecchia, P.~Kruchten, P.~Lago, and I.~Malavolta, ``{Building and
  evaluating a theory of architectural technical debt in software-intensive
  systems},'' \emph{Journal of Systems and Software}, vol. 176, p. 110925,
  2021.

\bibitem{Besker2022}
T.~Besker, A.~Martini, and J.~Bosch, ``{The use of incentives to promote
  technical debt management},'' \emph{Information and Software Technology},
  vol. 142, p. 106740, 2022.

\bibitem{avgeriou_technical_2023}
P.~Avgeriou, I.~Ozkaya, A.~Chatzigeorgiou, M.~Ciolkowski, N.~A. Ernst, R.~J.
  Koontz, E.~Poort, and F.~Shull, ``Technical {Debt} {Management}: {The} {Road}
  {Ahead} for {Successful} {Software} {Delivery},'' in \emph{2023 {IEEE}/{ACM}
  {International} {Conference} on {Software} {Engineering}: {Future} of
  {Software} {Engineering} ({ICSE}-{FoSE})}, May 2023, pp. 15--30.

\bibitem{Wiese2022}
M.~Wiese, P.~Rachow, M.~Riebisch, and J.~Schwarze, ``{Preventing technical debt
  with the TAP framework for Technical Debt Aware Management},''
  \emph{Information and Software Technology}, p. 106926, 2022.

\bibitem{AdditionalMaterial}
\BIBentryALTinterwordspacing
M.~Wiese, A.~Heinrichs, N.~Rusieshvili, and K.~B. Rebou{\c{c}}as~de Almeida,
  Rodrigo, ``Additional material for teh techdebt game - enabling discussions
  a,'' 2024. [Online]. Available: \url{https://doi.org/10.5281/zenodo.14205656}
\BIBentrySTDinterwordspacing

\bibitem{baker_problems_2003}
A.~Baker, E.~Navarro, and A.~van~der Hoek, ``Problems and {Programmers}: an
  educational software engineering card game,'' in \emph{25th {International}
  {Conference} on {Software} {Engineering}, 2003. {Proceedings}.}, May 2003,
  pp. 614--619, iSSN: 0270-5257.

\bibitem{fernandes_playscrum_2010}
J.~M. Fernandes and S.~M. Sousa, ``{PlayScrum} - {A} {Card} {Game} to {Learn}
  the {Scrum} {Agile} {Method},'' in \emph{2010 {Second} {International}
  {Conference} on {Games} and {Virtual} {Worlds} for {Serious} {Applications}},
  Mar. 2010, pp. 52--59.

\bibitem{heikkila_teaching_2016}
V.~T. Heikkilä, M.~Paasivaara, and C.~Lassenius, ``Teaching university
  students {Kanban} with a collaborative board game,'' in \emph{Proceedings of
  the 38th {International} {Conference} on {Software} {Engineering}
  {Companion}}, ser. {ICSE} '16.\hskip 1em plus 0.5em minus 0.4em\relax New
  York, NY, USA: Association for Computing Machinery, May 2016, pp. 471--480.

\bibitem{DeBoer2019}
R.~C. De~Boer, P.~Lago, R.~Verdecchia, and P.~Kruchten, ``{DecidArch} {V2}:
  {An} {Improved} {Game} to {Teach} {Architecture} {Design} {Decision}
  {Making},'' in \emph{Proceedings - 2019 {IEEE} {International} {Conference}
  on {Software} {Architecture} - {Companion}, {ICSA}-{C} 2019}, 2019, pp.
  153--157, iSBN: 9781728118765 Publisher: IEEE.

\bibitem{bass_software_2021}
L.~Bass, P.~Clements, and R.~Kazman, \emph{\BIBforeignlanguage{en}{Software
  {Architecture} in {Practice}: {Software} {Architect} {Practice}}},
  4th~ed.\hskip 1em plus 0.5em minus 0.4em\relax Addison-Wesley, 2021.

\bibitem{cervantes_smart_2016}
\BIBentryALTinterwordspacing
H.~Cervantes, S.~Haziyev, O.~Hrytsay, and R.~Kazman, ``Smart decisions: an
  architectural design game,'' in \emph{Proceedings of the 38th {International}
  {Conference} on {Software} {Engineering} {Companion}}, ser. {ICSE} '16.\hskip
  1em plus 0.5em minus 0.4em\relax New York, NY, USA: Association for Computing
  Machinery, May 2016, pp. 327--335. [Online]. Available:
  \url{https://dl.acm.org/doi/10.1145/2889160.2889184}
\BIBentrySTDinterwordspacing

\bibitem{Ganesh2014}
L.~Ganesh, ``Board game as a tool to teach software engineering concept -
  {Technical} debt,'' in \emph{Proceedings - {IEEE} 6th {International}
  {Conference} on {Technology} for {Education}, {T4E} 2014}, 2014, pp. 44--47,
  iSBN: 9781479964895 Publisher: IEEE.

\bibitem{TDGame1}
\BIBentryALTinterwordspacing
``{Technical Debt Game}.'' [Online]. Available:
  \url{https://www.tha.de/en/Computer-Science/THA-ias/Technical-Debt-Game.html}
\BIBentrySTDinterwordspacing

\bibitem{TDGame2}
\BIBentryALTinterwordspacing
``{Technical Debt Game -- for non-technical people}.'' [Online]. Available:
  \url{https://ri-level.de/technical-debt-game/}
\BIBentrySTDinterwordspacing

\bibitem{suits_what_1967}
B.~Suits, ``\BIBforeignlanguage{en}{What is a {Game}?}''
  \emph{\BIBforeignlanguage{en}{Philosophy of Science}}, vol.~34, no.~2, pp.
  148--156, Jun. 1967.

\bibitem{pivec_aspects_2003}
M.~Pivec, O.~Dziabenko, and I.~Schinnerl, ``Aspects of {GameBased} {Lerning},''
  in \emph{Proceedings of 3rd {International} {Conference} on {Knowledge}
  {Management}}, Graz, Austria, 2003.

\bibitem{plass_foundations_2015}
\BIBentryALTinterwordspacing
J.~L. Plass, B.~D. Homer, and C.~K. Kinzer,
  ``\BIBforeignlanguage{EN}{Foundations of {Game}-{Based} {Learning}},''
  \emph{\BIBforeignlanguage{EN}{Educational Psychologist}}, Oct. 2015,
  publisher: Routledge. [Online]. Available:
  \url{https://www.tandfonline.com/doi/abs/10.1080/00461520.2015.1122533}
\BIBentrySTDinterwordspacing

\bibitem{huizenga_teacher_2017}
J.~C. Huizenga, G.~T.~M. ten Dam, J.~M. Voogt, and W.~F. Admiraal, ``Teacher
  perceptions of the value of game-based learning in secondary education,''
  \emph{Computers \& Education}, vol. 110, pp. 105--115, Jul. 2017.

\bibitem{shi_game_2015}
Y.-R. Shi and J.-L. Shih, ``\BIBforeignlanguage{en}{Game {Factors} and
  {Game}-{Based} {Learning} {Design} {Model}},''
  \emph{\BIBforeignlanguage{en}{International Journal of Computer Games
  Technology}}, vol. 2015, no.~1, p. 549684, Jan. 2015, publisher: John Wiley
  \& Sons, Ltd.

\bibitem{garris_games_2002}
R.~Garris, R.~Ahlers, and J.~E. Driskell, ``\BIBforeignlanguage{en}{Games,
  {Motivation}, and {Learning}: {A} {Research} and {Practice} {Model}},''
  \emph{\BIBforeignlanguage{en}{Simulation \& Gaming}}, Dec. 2002, publisher:
  Sage PublicationsSage CA: Thousand Oaks, CA.

\bibitem{TechDebtGame_Tabletop}
\BIBentryALTinterwordspacing
``{Tabletop TechDebt Game on GitHub}.'' [Online]. Available:
  \url{https://github.com/TechDebtGame/TechDebtGame}
\BIBentrySTDinterwordspacing

\bibitem{TechDebtGame_Digital}
\BIBentryALTinterwordspacing
``{Digital TechDebt Game on Tabletopia}.'' [Online]. Available:
  \url{https://tabletopia.com/workshop/games/techdebts/1-7players/test}
\BIBentrySTDinterwordspacing

\bibitem{Freire2021a}
S.~S. Freire, N.~Rios, B.~P{\'{e}}rez, C.~Castellanos, D.~Correal,
  R.~Rama{\v{c}}, V.~Mandi{\'{c}}, N.~Tau{\v{s}}an, G.~L{\'{o}}pez, A.~Pacheco,
  D.~Falessi, M.~Mendon{\c{c}}a, C.~Izurieta, C.~Seaman, and R.~Sp{\'{i}}nola,
  ``{How Experience Impacts Practitioners' Perception of Causes and Effects of
  Technical Debt},'' in \emph{2021 IEEE/ACM 13th International Workshop on
  Cooperative and Human Aspects of Software Engineering (CHASE)}, 2021, pp.
  21--30.

\bibitem{Ernst2015}
N.~A. Ernst, S.~Bellomo, I.~Ozkaya, R.~L. Nord, and I.~Gorton, ``{Measure it?
  Manage it? Ignore it? Software practitioners and technical debt},''
  \emph{10th Joint Meeting of the European Software Engineering Conference and
  the ACM SIGSOFT Symposium on the Foundations of Software Engineering,
  ESEC/FSE 2015 - Proceedings}, pp. 50--60, 2015.

\bibitem{Besker2018b}
T.~Besker, J.~Bosch, A.~Martini, and J.~Bosch, ``{Technical debt cripples
  software developer productivity: A longitudinal study on developers' daily
  software development work},'' in \emph{Proceedings - International Conference
  on Software Engineering}.\hskip 1em plus 0.5em minus 0.4em\relax ACM, 2018,
  pp. 105--114.

\bibitem{Wohlin2012}
C.~Wohlin, P.~Runeson, M.~H{\"{o}}st, M.~C. Ohlsson, B.~Regnell, and
  A.~Wessl{\'{e}}n, \emph{{Experimentation in software engineering}}, 2012,
  vol. 9783642290442.

\bibitem{Martini2015a}
A.~Martini and J.~Bosch, ``{The Danger of Architectural Technical Debt:
  Contagious Debt and Vicious Circles},'' in \emph{Proceedings - 12th Working
  IEEE/IFIP Conference on Software Architecture, WICSA 2015}, 2015, pp. 1--10.

\bibitem{ramavc2022prevalence}
R.~Rama{\v{c}}, V.~Mandi{\'c}, N.~Tau{\v{s}}an, N.~Rios, S.~Freire,
  B.~P{\'e}rez, C.~Castellanos, D.~Correal, A.~Pacheco, G.~Lopez \emph{et~al.},
  ``Prevalence, common causes and effects of technical debt: Results from a
  family of surveys with the it industry,'' \emph{Journal of Systems and
  Software}, vol. 184, p. 111114, 2022.

\end{thebibliography}

\end{document}